\documentclass[sigconf, 10pt, anonymous=false, nonacm]{acmart}
\usepackage{algorithm}
\usepackage{algorithmic}
\usepackage{subfigure}
\usepackage{xspace}
\usepackage{comment}
\usepackage{soul}

\newcommand{\sn}{\textbf{\textit{mmMirror}}\xspace} 

\AtBeginDocument{%
  \providecommand\BibTeX{{%
    \normalfont B\kern-0.5em{\scshape i\kern-0.25em b}\kern-0.8em\TeX}}}

 \setcopyright{acmlicensed}
\copyrightyear{2024}
\acmYear{2024}
\acmDOI{XXXXXXX.XXXXXXX}

\acmConference[Conference acronym 'XX]{Make sure to enter the correct
  conference title from your rights confirmation email}{June 03--05,
  2018}{Woodstock, NY}
\renewcommand\footnotetextcopyrightpermission[1]{} 

\graphicspath{{./images/}}

\begin{document}

\title{\sn: Device Free mmWave Indoor NLoS Localization Using Van-Atta-Array IRS}




\author{Yihe Yan}
\affiliation{%
  \institution{University of New South Wales}
  \city{Sydney}
  \country{Australia}}
\email{yihe.yan@unsw.edu.au}

\author{Zhenguo Shi}
\affiliation{%
  \institution{University of New South Wales}
  \city{Sydney}
  \country{Australia}}
\email{zhenguo.shi@unsw.edu.au}

\author{Yanxiang Wang}
\affiliation{%
  \institution{University of New South Wales}
  \city{Sydney}
  \country{Australia}}
\email{yanxiang.wang@unsw.edu.au}

\author{Cheng Jiang}
\affiliation{%
  \institution{University of New South Wales}
  \city{Sydney}
  \country{Australia}}
\email{cheng.jiang1@student.unsw.edu.au}

\author{Chun Tung Chou}
\affiliation{%
  \institution{University of New South Wales}
  \city{Sydney}
  \country{Australia}}
\email{c.t.chou@unsw.edu.au}

\author{Wen Hu}
\affiliation{%
  \institution{University of New South Wales}
  \city{Sydney}
  \country{Australia}}
\email{wen.hu@unsw.edu.au}




\begin{abstract}

Industry 4.0 is revolutionizing manufacturing and logistics by integrating robots into shared human environments, such as factories, warehouses, and healthcare facilities. However, ensuring human safety in Non-Line-of-Sight (NLoS) scenarios, such as around corners, remains a critical challenge. Existing solutions, including vision-based and LiDAR systems, suffer from occlusions, lighting constraints, and privacy concerns, while RF-based systems face limitations in range and localization accuracy.



To address these limitations, we propose \textbf{\sn}, a novel system leveraging a Van Atta Array-based millimeter-wave (mmWave) reconfigurable intelligent reflecting surface (IRS) for precise, device-free NLoS localization. \sn integrates seamlessly with existing frequency-modulated continuous-wave (FMCW) radars and offers: (i) robust NLoS localization with centimeter-level accuracy at ranges up to 3 m, (ii) seamless uplink and downlink communication between radar and IRS, (iii) support for \textbf{multi-radar} and \textbf{multi-target} scenarios via dynamic beam steering, and (iv) reduced scanning latency through adaptive time slot allocation. Implemented using commodity 24 GHz radars and a PCB-based IRS prototype, \sn demonstrates its potential in enabling safe human-robot interactions in dynamic and complex environments.


\end{abstract}

\maketitle

\section{Introduction}
Ensuring the safety of humans and robots in shared environments is a critical challenge, particularly in Non-Line-of-Sight (NLoS) scenarios where objects or walls obstruct direct visibility. Robots operating in factories, warehouses, or other dynamic spaces often fail to detect humans or other obstacles around blind corners, leading to \textbf{collision risks}. This issue becomes increasingly important as robots are deployed in human-robot collaboration (HRC) environments.

\begin{figure}[htp]
    \centering
    \includegraphics[width=0.45\textwidth]{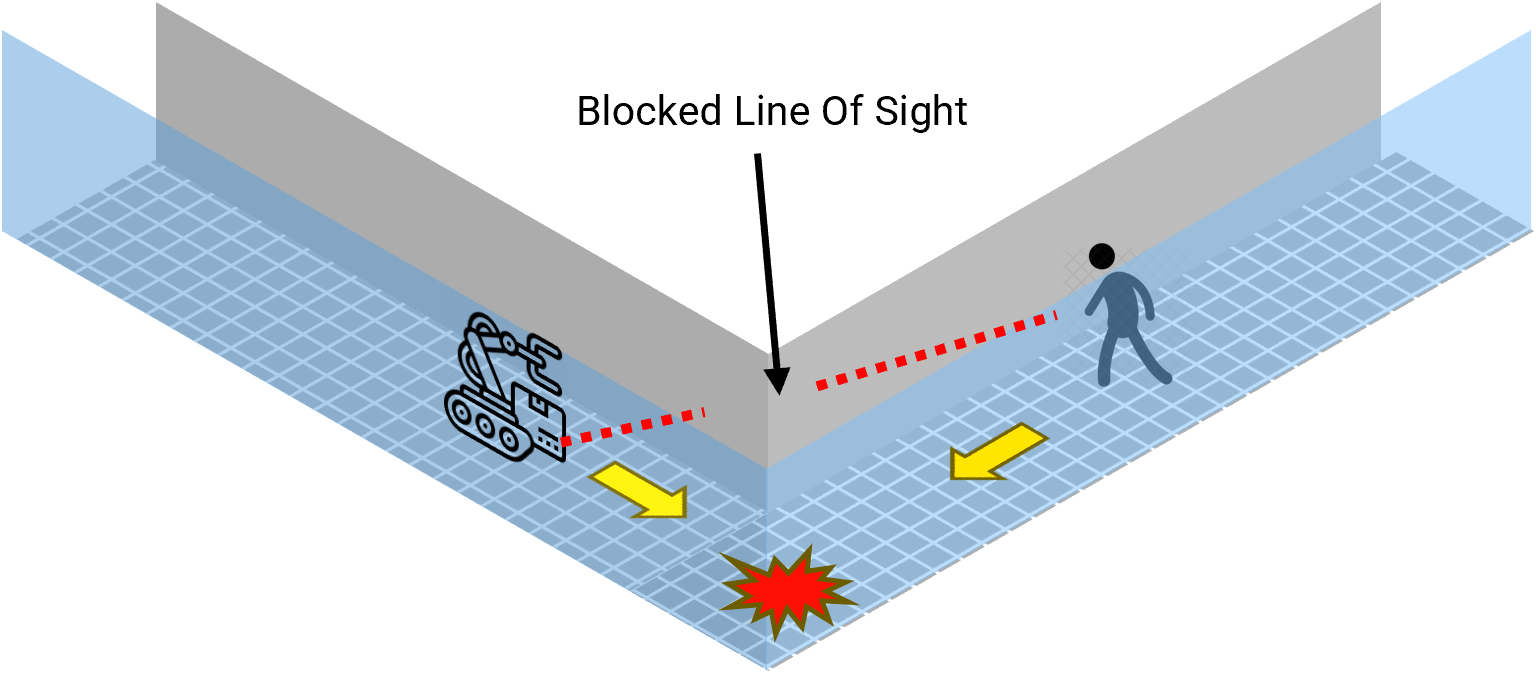}
    \caption{A typical scene of a collision. The robot and human are navigating around an L-shaped corridor with obstructed visibility. The lack of proper NLoS sensing results in a potential collision at the intersection point.}
    \label{fig:nlos_collision}
   \vspace{-0.3cm}
\end{figure}

Figure~\ref{fig:nlos_collision} illustrates a typical NLoS collision scenario. In this example, a robot and a human approach an L-shaped corridor, where the wall obstructs their visibility of each other. Without effective NLoS sensing, the robot cannot detect the human’s presence until it is too late, potentially causing a collision. Scenarios like this underscore the urgent need for robust, real-time localization systems that can operate effectively even in environments with occlusions.

Collisions in HRC environments \cite{han2021analysis} can cause harm to humans, disrupt operations, and lead to significant costs. In environments such as manufacturing plants or healthcare facilities, where robots are used for product transportation or human assistance \cite{antony2018food, robla2017working, okamura2010medical, goher2017assessment}, even a minor failure in localization can compromise safety and efficiency. Consider a scenario\cite{vslajpah2021effect} where a robot is moving at a speed of 0.8\,m/s while a human is walking at 1\,m/s. If both the robot and the human start 3\,m away from the corridor, a collision will occur in just 3.3\,s. Effective NLoS sensing is, therefore, vital for enabling robots to navigate safely and interact seamlessly with their surroundings.

Several sensing techniques have been developed to tackle NLoS challenges. Vision-based systems \cite{naser2019infrastructure, yedidia2019using, lindell2019wave, ye2024plug, li2023nlost, velten2012recovering} provide high-resolution localization but are hindered by sensitivity to lighting conditions and privacy concerns. LiDAR-based solutions \cite{ayyalasomayajula2020deep, malik2024transient} offer robust detection in some scenarios but are costly and limited by hardware complexity. RF-based technologies \cite{yue2022cornerradar, li2020wiborder, zetik2015looking}, including Wi-Fi \cite{gunia2023analysis, adib2013see, yao2024wiprofile}, Bluetooth \cite{yu2021novel}, and RFID \cite{bouet2008rfid}, use signal reflections to locate targets but suffer from low accuracy and range limitations. While recent advancements in \textbf{millimeter-wave (mmWave)} technologies improve localization accuracy, most methods rely heavily on Line-of-Sight (LoS) to the target. This dependence renders them ineffective in environments with occlusions caused by walls or furniture. 

Existing NLoS solutions face significant limitations. Passive methods \cite{xu2023leveraging, woodford2022mosaic, wei2021nonline} depend on favorable surface properties for reflections and often require additional hardware like LiDAR. Active approaches, such as those utilizing \textbf{multiple} retro-reflective tags \cite{bae2024supersight, wang2013dude} or metasurfaces \cite{woodford2023metasight, chen2016review}, demand target-side modifications or complex hardware designs, which are impractical in dynamic and human-shared environments that require adaptability and 
communications between the radar and the metasurface without out-of-band channels. 
Furthermore, some of them\cite{okubo2024integrated, bae2024supersight, wang2013dude} require the target to be localized to carry retro-reflective tags, which is inconvenient.

Additionally, a recent approach~\cite{Dodds2024Around} demonstrated NLoS mmWave imaging, achieving high-resolution imaging of approximately 0.02\,m around corners without relying on strong environmental assumptions. However, this method primarily focuses on generating point clouds in NLoS environments and requires CUDA for processing 3D radar heatmaps. This reliance on CUDA poses a limitation, as not all robots support CUDA, and the approach does not facilitate real-time target detection.

Therefore, existing solutions are incapable of providing both \textbf{device-free precise localization} and \textbf{real-time adaptability} in scenarios involving multiple targets and dynamic obstacles.


To address these challenges, we propose \textbf{\sn}, a novel device-free NLoS localization system that leverages a \textbf{Van Atta Array (VAA)-based Intelligent Reflecting Surface (IRS)} for precise mmWave localization that does not require the target to carry a tag or device. To the best of our knowledge, no prior work has fully integrated a VAA-based IRS system for NLoS localization that \textbf{combines sensing, communication, and adaptive beam control into a unified framework}. Our work is the \textbf{first to design, implement, and evaluate such a system}. Unlike exhaustive scanning methods, our adaptive sensing mechanism prioritizes critical directions, \textbf{significantly reducing latency — a crucial factor for real-time collision detection} — while maintaining accuracy. \sn achieves robust localization by a novel \textbf{reconfigurable VAA-based IRS} to enable stable, adjustable reflection paths, via changing the length of transmission line between paired VAA antennas in real-time, at a lower cost compared to traditional metasurfaces ~\cite{kim2024nr}. We further introduce an \textbf{amplitude modulation-based antenna encoding technique} to facilitate seamless radar-IRS communication using frequency-modulated continuous-wave (FMCW) signals. Additionally, we develop an \textbf{adaptive time slot scheduling algorithm} to optimize sensing by focusing on Areas of Interest (AoI), reducing scanning latency, and improving signal-to-noise ratio (SNR). The key contributions of this paper are:
\begin{itemize}
\vspace{-0.1cm}
    \item We propose a novel device-free mmWave-based NLoS localization system that achieves centimeter-level accuracy with a single mmWave radar and a VAA-based IRS.
    \item An innovative antenna encoding method has been proposed which utilizes two TX antennas and dynamically adjusts transmission power, seamlessly integrating radar sensing and communication while supporting multiple radars. Additionally, we introduce a time slot allocation algorithm that reduces scanning latency by up to 42\% for multiple targets and 75\% for a single target, significantly enhancing localization efficiency.
    \item We conduct extensive experiments that demonstrate \sn's ability to achieve a localization error of 8.93\,cm at a 3\,m range in NLoS environments, highlighting its practicality for human-robot interaction.
    \vspace{-0.3cm}
\end{itemize}

\section{Background Knowledge}

Table~\ref{tab:symbols} provides a summary of the mathematical symbols used throughout this paper.

\begin{table}[ht]
\centering
\caption{Mathematical Symbols and Their Meanings}
\resizebox{\columnwidth}{!}{%
\begin{tabular}{|c|l|}
\hline
\textbf{Symbol} & \textbf{Meaning} \\ \hline
$A$ & Amplitude of the transmitted signal \\ \hline
$f_0$ & Starting frequency of the chirp \\ \hline
$\beta$ & Chirp slope, calculated as $\beta = \frac{B}{T_{\text{chirp}}}$ \\ \hline
$\tau$ & Round-trip time related to target range $R$, $\tau = \frac{2R}{c}$ \\ \hline
$c$ & Speed of light \\ \hline
$R$ & Range of the target \\ \hline
$f_b$ & Beat frequency, $f_b = \beta \tau = \beta \frac{2R}{c}$ \\ \hline
$v$ & Velocity of the target \\ \hline
$\lambda$ & Wavelength of the signal \\ \hline
$\phi_n$ & Phase delay for the $n$-th antenna element \\ \hline
$\delta$ & Phase gradient for beam steering \\ \hline
$\epsilon_r$ & Relative permittivity of the substrate \\ \hline
$\lambda_g$ & Guided wavelength in the microstrip line \\ \hline
$D_{RS}$ & Distance between radar and IRS \\ \hline
$D_{ST}$ & Distance between IRS and target \\ \hline
$\phi$ & Angle of arrival (AoA) \\ \hline
$\alpha_i$ & Reflection angle \\ \hline
$x_R, y_R$ & Coordinates of the radar \\ \hline
$x_S, y_S$ & Coordinates of the IRS \\ \hline
$x_T, y_T$ & Coordinates of the target \\ \hline
\end{tabular}%
}
\label{tab:symbols}
\end{table}

\subsection{FMCW Signal Model}

The transmitted signal in an FMCW system~\cite{adib2015multi, orth2019novel} is a chirp, where the frequency linearly increases (or decreases) over time. The transmitted signal $s_{\text{tx}}(t)$ is expressed as:
\begin{eqnarray}
s_{\rm tx}(t) & = & 
A \cos\left(2 \pi \left( f_0 t + \frac{\beta}{2} t^2 \right) \right)
\label{eq:fmcw:chirp}
\end{eqnarray}
where $A$ is the amplitude, $f_0$ is the starting frequency, $t$ is time, and $\beta = \frac{B}{T_{\text{chirp}}}$ is the chirp slope where \( B \) is the bandwidth of the chirp and \( T_{\text{chirp}} \) is the chirp duration.

The received signal is delayed by the round-trip time $\tau$, which is related to the target range $R$ by $
\tau = \frac{2R}{c},
$
where \( c \) is the speed of light. 

In FMCW, the transmitted signal is mixed with the received signal to produce a beat signal
with frequency $f_b$ 
given by 
$
f_b = \beta \tau = \beta \cdot \frac{2R}{c}.
$

For a moving target, the Doppler shift $f_d$ is introduced, and the total beat frequency becomes
$
f_{\text{total}} = f_b + f_d,
$
where 
$
f_d = \frac{2v}{\lambda}, 
$
where $\lambda$ is the wavelength and $v$ is the target's velocity. By separating $f_b$ and $f_d$, the system estimates both the target's range $R$ and velocity $v$ .

\subsection{Van Atta Array}

Proposed in the 1960s~\cite{sharp1960van}, the VAA employs antenna reciprocity to achieve retro-reflection. A VAA consists of paired antennas configured to cancel unwanted signals and reinforce desired ones through phase manipulation. This enables tasks such as beam steering without physically moving the antennas.

Let the incoming wave be $E_{\text{in}}(t)$:
\[
E_{\text{in}}(t) = E_0 \cos(2 \pi f t + \phi),
\]
where $E_0$ is amplitude, $f$ is frequency, and $\phi$ is phase. The $n$-th antenna element receives this signal with a phase delay $\phi_n$, expressed as:
\[
E_n(t) = E_0 \cos(2 \pi f t + \phi_n).
\]

In a VAA, antennas are paired such that the $n$-th antenna's received signal is re-radiated with an inverted phase by the $-n$-th antenna, reinforcing retro-reflection. This results in improved SNR, range, and resolution.

\subsection{Non-Retroreflective Van Atta Array}
\label{subsec:NVAA}

The Non-Retroreflective VAA~\cite{ang2018passive} customizes antenna pairings to control reflection direction. For example, the $n$-th antenna may be connected to the $m$-th antenna ($m \neq -n$), producing an outgoing signal:
\[
E_{\text{out}}(t) = E_0 \cos(2 \pi f t + \phi_{\text{out}}),
\]
where $\phi_{\text{out}}$ depends on the pairing and desired reflection angle. This design enables beam steering without active electronics, suitable for applications in \textbf{low power} radar and communications.

\subsection{Reflection-Based NLoS Localization}

\label{section:reflection_based_localization}

\begin{figure}[!htp]
    \centering
    \includegraphics[width=0.25\textwidth]{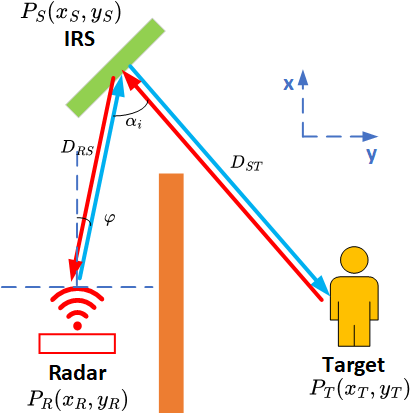}
    \caption{A Typical NLoS Environment Layout}
    \label{fig:nlos_layout}
    \vspace{-0.5cm}
\end{figure}

Figure~\ref{fig:nlos_layout} shows a typical NLoS scenario with a radar at $P_R(x_R, y_R)$, an IRS at $P_S(x_S, y_S)$, and a target at $P_T(x_T, y_T)$. An (orange) obstacle blocks the direct LoS path. The IRS redirects the incident radar signal by a \textbf{reflection angle} $\alpha_i$ along the path $P_R \rightarrow P_S \rightarrow P_T$ and back. By using the radar-to-IRS distance $D_{RS}$ and the angle of arrival (AoA) $\varphi$, we can determine the IRS's position as:
\begin{eqnarray}
x_S & = & x_R + D_{RS} \cos(\varphi) 
\label{eq:loc:irs_x}
\\
y_S & = & y_R + D_{RS} \sin(\varphi), 
\label{eq:loc:irs_y}
\end{eqnarray}
The target's position is then calculated as:
\begin{eqnarray}
    x_T & = & x_S - D_{ST}\cos\left( \alpha_i -  \varphi\right),
\label{eq:pos_x} \\
   y_T & = &  y_S + D_{ST} \sin\left( \alpha_i -  \varphi\right),
\label{eq:pos_y}
\end{eqnarray}
where $D_{ST}$ is the IRS-to-target distance. Note that, unlike existing NLoS method ~\cite{ woodford2022mosaic} which uses LiDAR to determine reflector orientation, the above methods computes the target location without having to know the orientation of the IRS. 
Our novel IRS design addresses this by enabling both reflection and communication, which will be discussed next.

\section{\sn System Design}

This section details the design of the \sn system, beginning with an overview and proceeding through its key components: hardware design, IRS-radar communication, and localization procedure.

\subsection{\sn Overview}

\begin{figure}[t]
    \centering
    \includegraphics[width=0.5\textwidth]{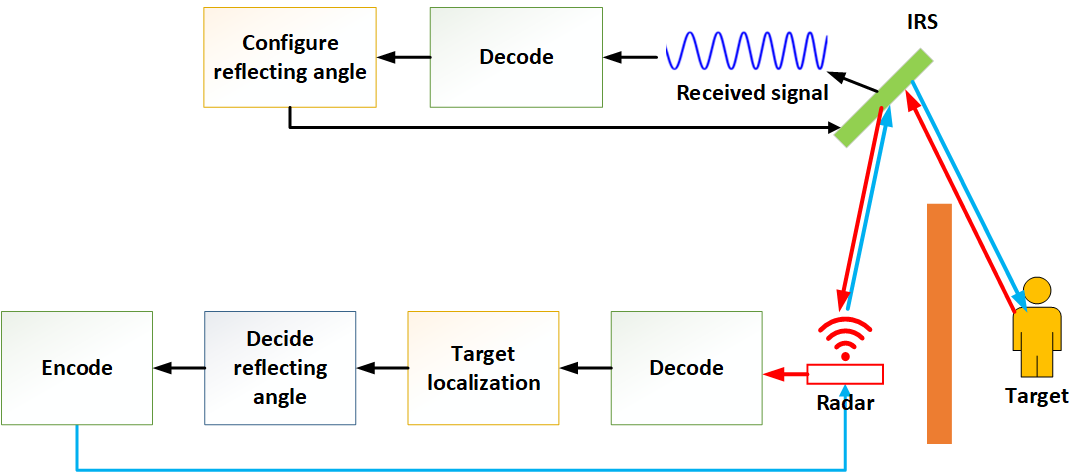}
    \caption{\sn System Overview.}
    \label{fig:nlos_working_overview}
    \vspace{-0.3cm}
\end{figure}

\sn is a novel mmWave-based system tailored for precise localization in NLoS scenarios. Figure~\ref{fig:nlos_working_overview} outlines the \sn workflow, which comprises the key steps as follows. First, the IRS operates in the retro-reflective mode and communicates its presence to the radars by encoding the received signal from the radars before retro-reflecting it. The radars can use the retro-reflected signal from the IRS to localise the IRS. Second, using antenna encoding, the radar transmits FMCW chirps to define AoI for localization. The IRS decodes these signals and dynamically configures its reflection angle based on the received information. Third, the radar detects potential targets and determines optimal scanning areas. To enhance SNR, \sn dynamically allocates reflection times to different angles and prioritizes scanning angles containing active targets. This approach minimizes latency and improves localization precision. Finally, by leveraging position and angle information provided by the IRS, \sn refines AoI for subsequent scans. This iterative process ensures efficient tracking of moving targets.

\subsection{\sn Hardware Design}
\label{subsec:hardwareDesign}

\sn's hardware leverages a non-retro-reflective VAA-based IRS to reflect 
received radio energy to different angles via adjusting the length of
transmission line between paired antennas as discussed in Section~\ref{subsec:NVAA} earlier.

\begin{figure}[t]
    \centering
    \includegraphics[width=0.4\textwidth]{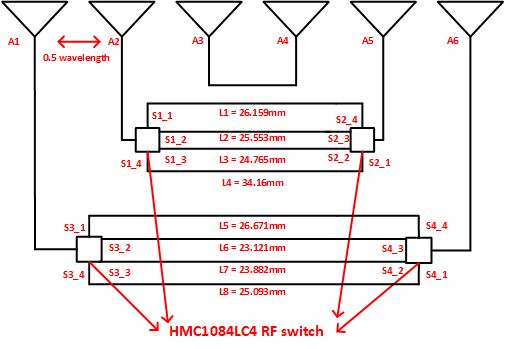}
    \caption{4-Way Reflective VAA Antenna Design.}
    \label{fig:antenna}
    \vspace{-3mm}
\end{figure}

\begin{table}[htp]
\centering
\caption{Mapping of Reflection Angles to Transmission Lines and Switch Configurations
for the Reflective VAA Antenna Design in Figure~\ref{fig:antenna}.}
\scalebox{0.75}{
\begin{tabular}{|c|c|c|}
\hline
\textbf{Reflection Angle} & \textbf{Transmission Line Selected} & \textbf{Switch Enabled} \\ \hline
30° & L3, L5 & S1\_3, S2\_2, S3\_1, S4\_4 \\ \hline
45° & L2, L8 & S1\_2, S2\_3, S3\_4, S4\_1 \\ \hline
60° & L1, L7 & S1\_1, S2\_4, S3\_3, S4\_2 \\ \hline
75° & L4, L6 & S1\_4, S2\_1, S3\_2, S4\_3 \\ \hline
\end{tabular}
}

\label{table:reflection_angles}
\end{table}

Figure~\ref{fig:antenna} shows an example hardware prototype that operates at $f$ = 24 GHz and features eight transmission lines with different lengths connected to six patch antennas spacing by half wavelength of the radio signal to enable \textit{beam steering}. To maximize NLoS coverage and minimize errors from angular mismatch, we designed a system supporting four distinct reflection angles with a discrete 15° offset (30°, 45°, 60°, and 75°). This reduces the maximum angular mismatch error to 10.4\,cm. The design integrates multiple transmission line sets with four switches, adjusting the lengths between paired antennas to optimize scanning in NLoS scenarios.

\begin{figure*}[htp]
    \centering
    \includegraphics[width=0.98\textwidth]{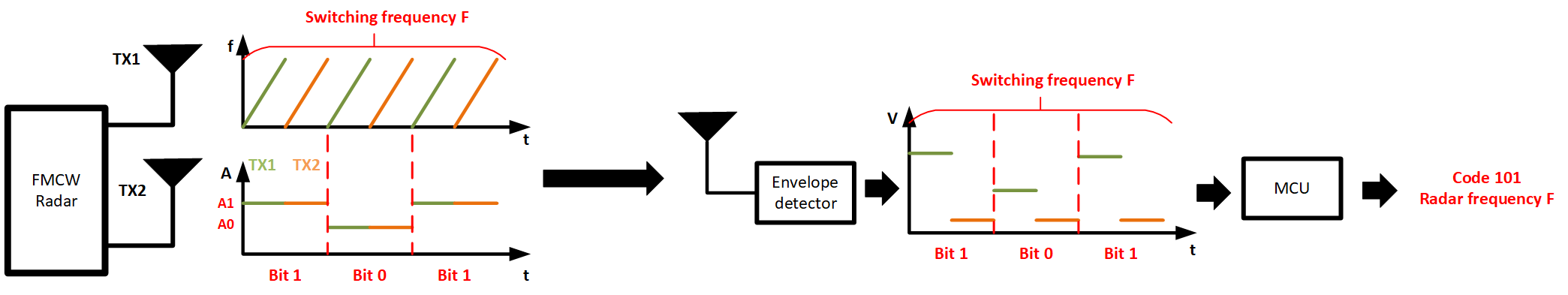}
    \caption{Radar to IRS communication in \sn.}
    \label{fig:encoding_overview}
\end{figure*}


In addition, the series-fed patch array elements, which provide high gain, maximize the energy of radar signals reflected into the NLoS area. The patch antennas employ an insert-fed technique, ensuring compact size and high reflection coefficient (i.e., \( S_{11} \)) performance.

Beyond two-way reflection, our hardware design also supports a "conventional" \textbf{retro-reflection VAA mode} by using the OFF state of the switches, where no transmission lines are connected. This flexible and scalable design balances performance and cost, making \sn suitable for real-world, device-free NLoS sensing applications.

\subsection{IRS-Radar Communication}
\label{subsec:IRS-RadarCommunication}
This section explains how \sn implements bidirectional communication between the radar and the IRS. In \sn, the radar instructs the IRS on which reflection angles to use, and the IRS broadcasts the reflection angle it is currently using to all radars.

\textbf{IRS to Radar Communication:} The IRS employs On-Off Keying (OOK) modulation to encode information. Specifically, the IRS modulates the radar's radio signal by toggling its switches in retro-reflection mode: the ON state represents Symbol "1," and the OFF state represents Symbol "0," as inspired by Milimetro~\cite{soltanaghaei2021millimetro}. The radar demodulates the retro-reflected signal by analyzing variations in reflection strength, enabling it to simultaneously localize the IRS (using Eq. (\ref{eq:loc:irs_x})--(\ref{eq:loc:irs_y}) after estimating range \( D_{RS} \) and AoA \( \varphi \)) and decode transmitted data.




\textbf{Radar to IRS Communication:} 

Figure~\ref{fig:encoding_overview} presents an overview of the communication from the radar to the IRS. The key components are a novel antenna encoding method at the radar (which exploits multiple transmission antennas that are commonly available in the Commercially-Off-The-Shelf or COTS mmWave radars), and an envelope detector and a MCU for bit detection at the IRS.

The radar encodes the data stream as binary bits using a combination of amplitude modulation and an antenna switching frequency $F$ where $F \ll f_0$ with $f_0$ being the chirp carrier frequency.  Each radar uses two of its transmission antennas TX1 and TX2.  The duration of one bit is $\frac{2}{F}$ and is divided two equal halves where TX1 and TX2 transmit in each half as illustrated in Figure~\ref{fig:encoding_overview}. When a Bit 1 (resp. Bit 0) is transmitted, the radar uses a higher amplitude $A_1$ (resp. lower amplitude $A_0$) for the chirp signal in Eq. (\ref{eq:fmcw:chirp}), see Figure~\ref{fig:encoding_overview}. We remark that each radar uses a unique antenna switching frequency $F$ and we will discuss this under \textsl{multiple radar support} later in this section.  

In order for the IRS to distinguish between Bit 0 and Bit 1 sent by the radar, it only needs to obtain the signal amplitude during the bit duration. This means the typical de-chirping circuitry consisting of an oscillator, multiplier, low-pass filter is \textsl{not} necessary. We have chosen to use an envelope detector to obtain the received signal amplitude. 

The envelope detector circuit, comprising a diode, capacitor, and resistor.  When an amplitude-modulated signal passes through the diode, the diode rectifies the signal, allowing only the positive half-cycles to pass through. The capacitor then charges to the peak amplitude of the signal, creating a smoothed waveform that represents the envelope of the input. The resistor discharges the capacitor during the low-amplitude portions of the signal, which allows the circuit to follow the variations in the amplitude over time, accurately producing the envelope. The output voltage \( V_{\text{out}} \) of an envelope detector depends on the peak amplitude \( A_{\text{in}} \) of the received signal and is approximately given by:
\[
V_{\text{out}} \approx A_{\text{in}} - V_{\text{diode}}.
\]
where $V_{\text{diode}}$ is the voltage across the diode. 

\begin{figure}[htp]
    \vspace{-3mm}
    \centering
    \includegraphics[width=0.48\textwidth]{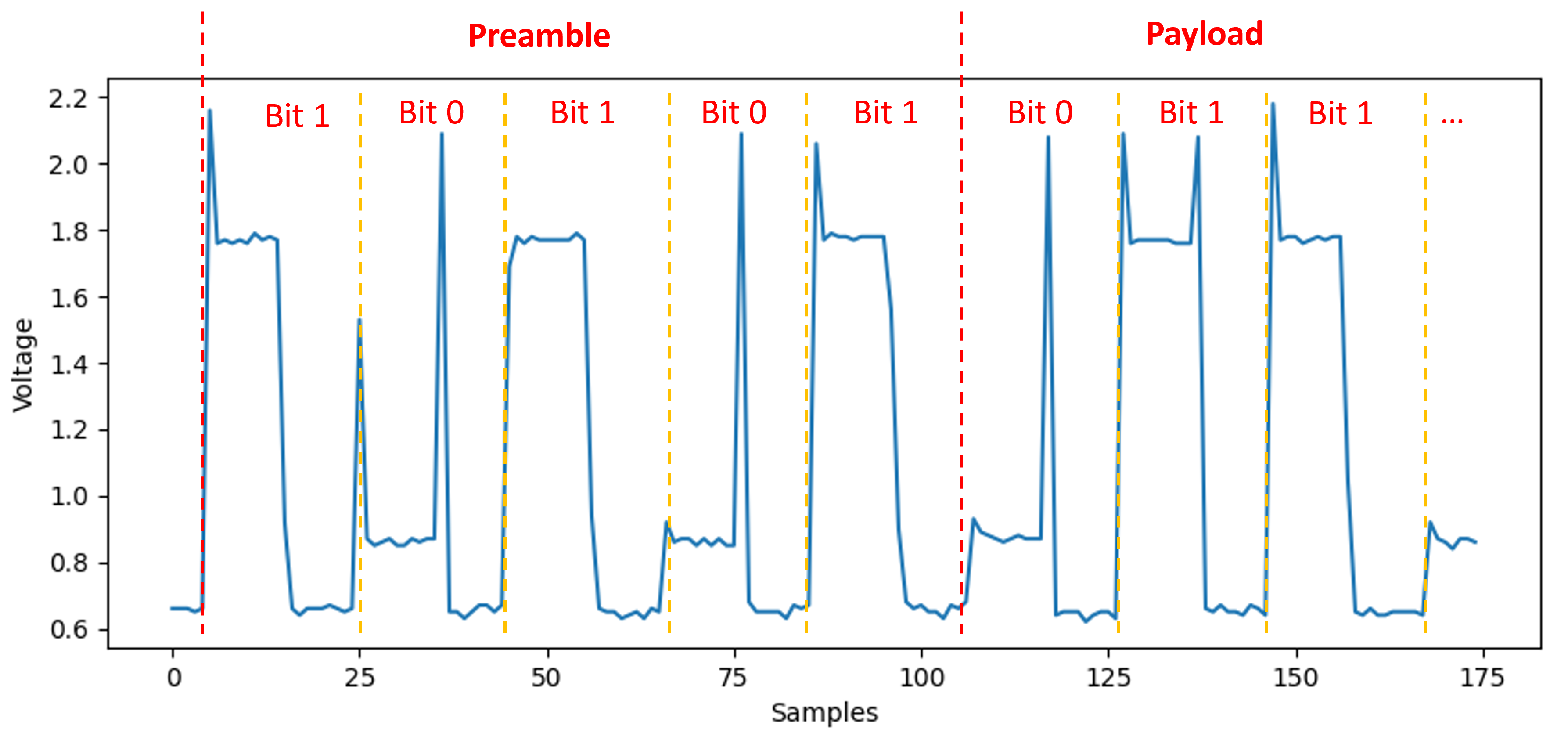}
    \caption{A \sn Antenna Encoding Raw Signal Example}
    \label{fig:data_example}
    \vspace{-3mm}
\end{figure}

Figure~\ref{fig:encoding_overview} illustrates the output of the envelope detector and Figure \ref{fig:data_example} shows the output from the \sn prototype. There are two features that we want to point out. First, the received amplitudes from TX1 and TX2 transmissions are different. This is because then distances from TX1 and TX2 to the IRS are different. We will use this to obtain information on the switching frequency $F$ to allow multiple access from different radars; this will be discussed under \textsl{multiple radar support} later on. Second, the received amplitude for Bit 1 is higher than that of Bit 0. For detection, we decide for Bit 0 if its received amplitude is smaller than 0.6 times the received amplitude for Bit 1.

Lastly the envelop detector output in Figure~\ref{fig:data_example} illustrates the packet structure we use. A packet consists of 2 bits for packet identification, 3 bits for synchronization and a 6-bit data field for payload. \newline


\textbf{Radar to IRS communication Data Rate:} 
The radar symbol is defined by the number of TX antenna repetitions, \( N_r \), and the antenna switching frequency, \( F \). Consequently, the achievable data rate can be calculated using the following equation:

\[
\text{data\_rate} = \frac{F}{2N_r}
\]
where the number 2 comes from the fact that two transmission antennas take turns to transmit in the duration of a bit, see Figure~\ref{fig:encoding_overview}. 

To increase the data rate, we can either reduce the number of TX antenna repetitions or increase the antenna switching frequency. For example, with only one TX antenna repetition and an antenna switching frequency of 3,906.25 Hz (corresponding to a chirp period of \( 256 \, \mu\text{s} \)), a data rate of 1,953.125 bps can be achieved in theory. However, due to hardware limitations of our prototype, the maximum configurable antenna switching frequency is restricted to 20 Hz. Despite this limitation, the achievable data rate is sufficient for our purpose, i.e., for the radar to communicate the desired reflection angles, and time, i.e., sensing slots, dynamically for the IRS.\newline

\begin{figure}
	\centering
    \includegraphics[width=0.45\textwidth]{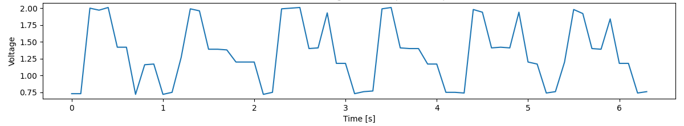}
    \centerline{(a)}
	\\
    \centering
    \includegraphics[width=0.45\textwidth]{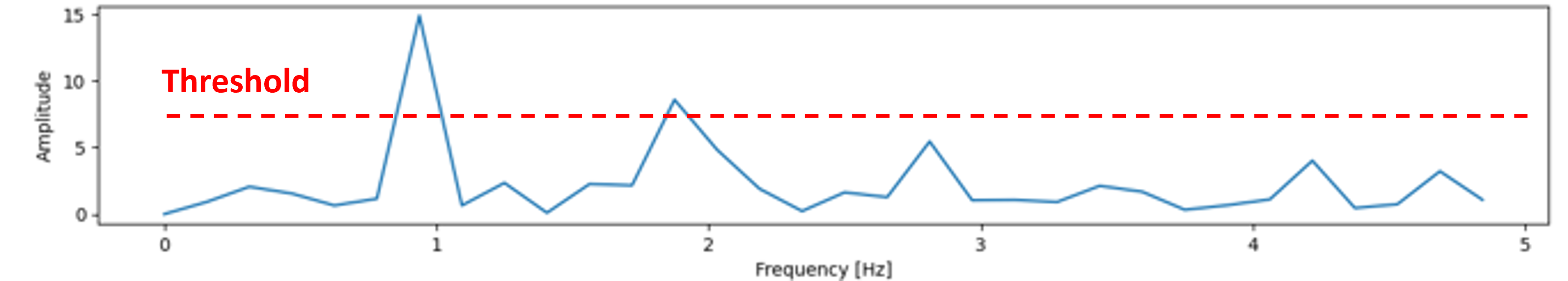}
    \centerline{(b)}
    \\
    \centering
    \includegraphics[width=0.45\textwidth]{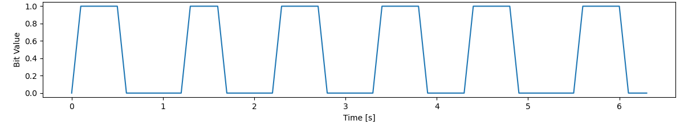}
    \centerline{(c)}
    \\
    \centering
    \includegraphics[width=0.45\textwidth]{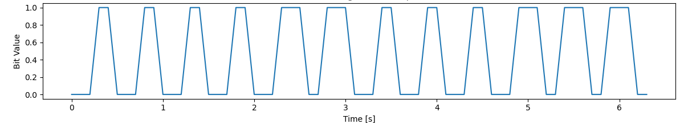}
    \centerline{(d)}
	
    \caption{(a) Raw signal output from the envelope detector with two radars transmitting coded 1010 sequences at 1 Hz and 2 Hz. (b) FFT result of the raw signal, highlighting the frequency components corresponding to the transmitted sequences. Decoded signal for 1 Hz (c) and 2 Hz (d) transmission after applying a band-pass filter.}
	\label{fig:multiple_radar_example}
    \vspace{-0.3cm}
\end{figure}

\textbf{Multiple radar support:} 
Nowadays, most mmWave radars have two or more TX antennas \cite{radar, radar1, radar2, radar3}. To fully leverage this advantage, we designed an antenna encoding scheme that utilizes two antennas and allocating a unique antenna switching frequency \( F \) to each radar. For this multiple access problem, the IRS needs to determine the number of radars and their identities (by finding out the $F$ they use to transmit), and to decode the radars' simultaneous transmissions. 

To achieve these goals, the IRS applies Fast Fourier Transform (FFT) on the signal after the envelope detector where each antenna switching frequency $F$ will result in a peak in the FFT. The IRS uses a threshold on the FFT amplitudes to determine the number of radars and the frequencies at the peaks will give the identities of the radars.  The threshold is empirically set to be the half of the first peak. Finally, the IRS separates the signals from different radars using band pass filters. An example with two radars of switching frequencies 1 Hz and 2 Hz respectively, is shown in Figure \ref{fig:multiple_radar_example}. 

The accuracy of the proposed method depends on the MCU sampling frequency where higher sampling frequency results in more accurate detection. However, in order to lower the power consumption, we limit the MCU sampling frequency to 64 Hz, which is sufficient to distinguish 2 radars as it is less frequent for a group of robots to approaching an intersection point causing potential collision.

In summary, this communication framework is optimized for minimal hardware complexity and supports scenarios involving multiple radars. 


\subsection{Localization Procedure}

\begin{figure}[htbp]
    \begin{minipage}[t]{0.5\linewidth}
        \centering
        \includegraphics[width=0.97\textwidth]{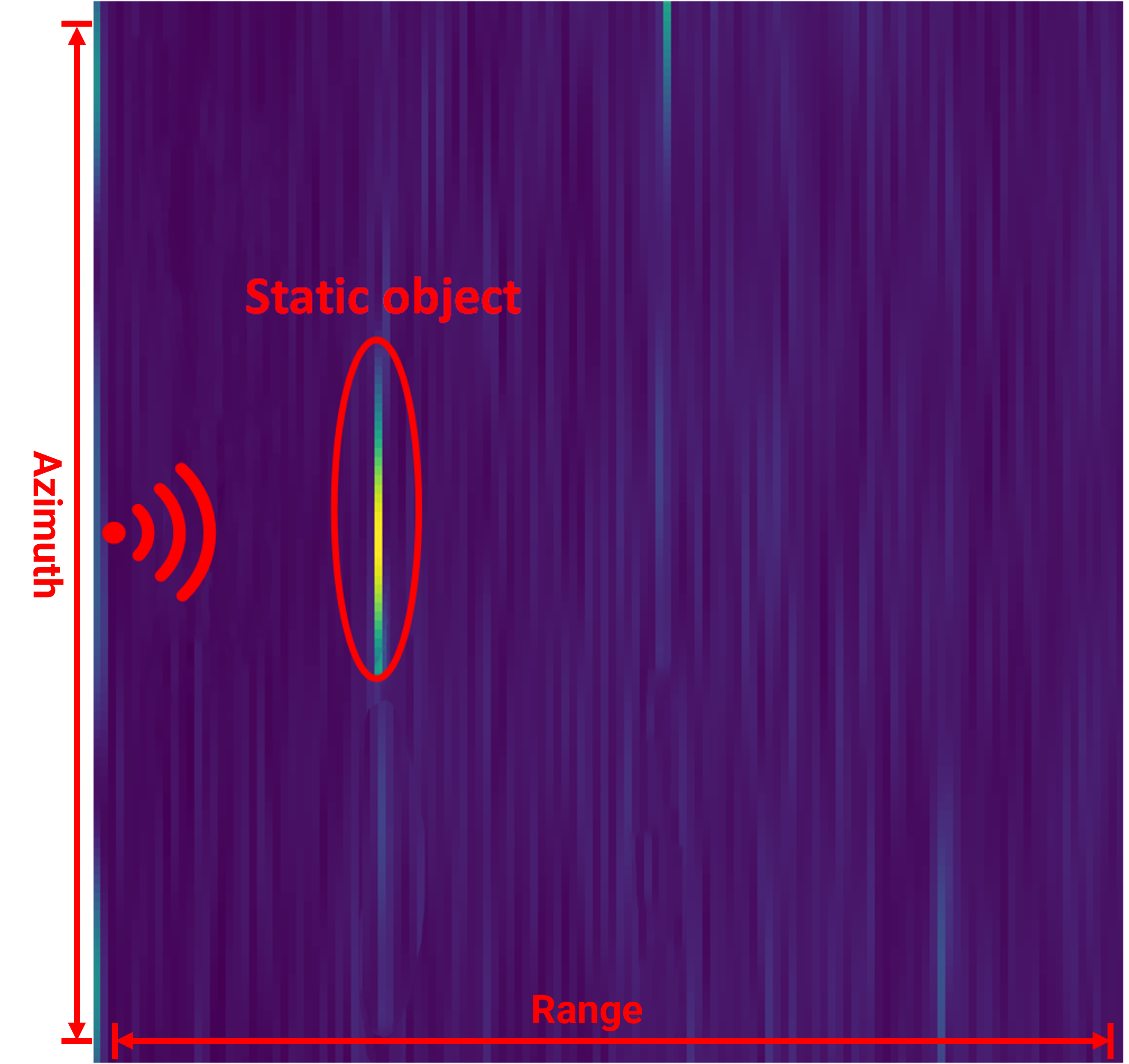}
        \centerline{(a)}
    \end{minipage}%
    \begin{minipage}[t]{0.5\linewidth}
        \centering
        \includegraphics[width=0.9\textwidth]{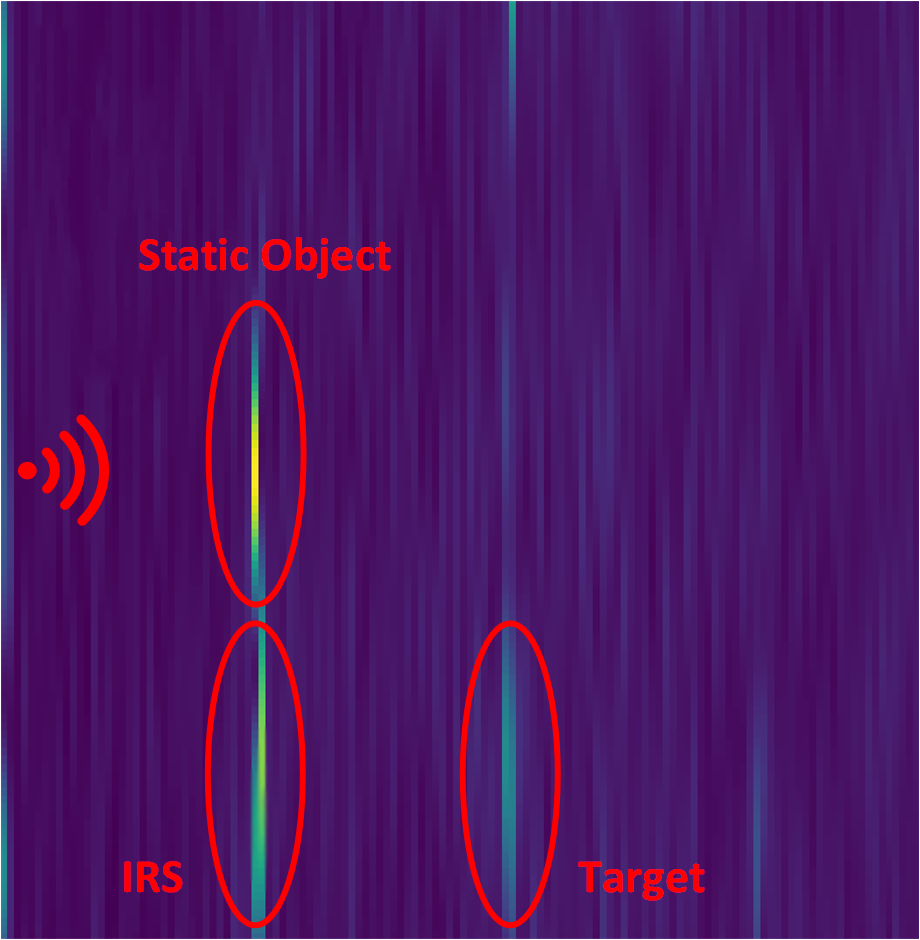}
        \centerline{(b)}
    \end{minipage}
    \caption{(a) IRS in retro-reflection mode,  IRS is in OFF status. (b) IRS in ON status and reflection mode. }
    \label{fig:vaa_mode}
\end{figure}


\begin{algorithm}
\caption{\sn IRS Working Procedure}
\label{alg:irs_working}
\begin{algorithmic}[1]
\STATE \textbf{Start:} IRS broadcasts ID.
\WHILE{TRUE}
    \STATE IRS takes ADC samples.
    \IF {Radar signal exists}
        \STATE IRS decodes the signal.
        \IF {Signal contains AoI (Area of Interest)}
            \STATE Use AoI set from radar.
        \ELSE
            \STATE Use all reflecting angles as AoI.
        \ENDIF
        \FOR {Each angle $\alpha$ in AoI}
            \STATE Broadcast the angle $\alpha$ in retro-reflection mode.
            \STATE Switch the transmission lines to use  reflection angle $\alpha$ in reflection mode.
        \ENDFOR
    \ELSE
        \STATE Repeat ADC sampling.
    \ENDIF
\ENDWHILE
\end{algorithmic}
\end{algorithm}


Figure~\ref{fig:vaa_mode} shows a one-shot radar image with the IRS turned off (a) and on (b), with noise filtered out using a relative thresholding method. From this figure, we observe that the reflected energy from the wall is \(-76\) dB, compared to \(-63\) dB, reflected from IRS, making it too weak to be suitable for real-time NLoS sensing.

Since the radar initially lacks knowledge of the environment, the retro-reflection mode (see Section~\ref{subsec:hardwareDesign}) enables the radar to localize the position of the IRS and distinguish it from static objects, as shown in Figure \ref{fig:vaa_mode}. The figure also shows that targets in NLoS area (and IRS) appear in retro-reflection mode Figure~\ref{fig:vaa_mode}(b) only but not in reflection mode Figure~\ref{fig:vaa_mode}(a), which shows that \sn can distinguish NLoS targets and LoS targets by alternating retro-reflection and reflection modes. During this period, the IRS uses OOK modulation discussed in Section~\ref{subsec:IRS-RadarCommunication} earlier to broadcast its unique ID, which provides potential support for scenarios with multiple IRS devices.

Algorithm~\ref{alg:irs_working} outlines the \sn IRS working procedure, where the IRS operates as a master device to enable dynamic NLoS localization. The process begins with the IRS broadcasting its unique ID (Line 1) to allow radars in the vicinity to identify and localize the IRS (see Algorithm~\ref{alg:radar_working} later for the details). The IRS then enters an infinite loop (Line 2) to continuously monitor for radar signals by taking Analog-to-Digital Converter (ADC) samples as the input (Line 3). If a radar signal is detected by IRS' \textit{envelop detector}, the IRS decodes the signal (Line 5) withe the algorithm discussed in Section~\ref{subsec:IRS-RadarCommunication} earlier. Based on the decoded signal, the IRS either uses the AoI provided by the radar (see Section~\ref{subsec:resourceAllocation} below for the details) or defaults to using all reflection angles as the AoI (Lines 7 - 9). For each angle in the AoI, the IRS broadcasts that angle in retro-reflection mode (Line 11) and switches to the corresponding transmission line (e.g., in Table~\ref{table:reflection_angles}) so that it reflects using that angle (Line 12). If no radar signal is detected, the IRS repeats ADC sampling (Line 14) and to continue monitoring. This process enables the IRS to dynamically adjust its behavior based on radar inputs, ensuring efficient and precise localization in NLoS scenarios. Continuous operation (Line 16) ensures the system remains adaptive and responsive in real time.

\begin{algorithm}
\caption{\sn Radar Working Procedure}
\label{alg:radar_working}
\begin{algorithmic}[1]
\STATE \textbf{Start:} Radar sending out FMCW chirps.
\WHILE{TRUE}
    \STATE Radar receives reflected signal.
    \IF {IRS OOK modulation exists}
        \STATE Radar decodes the IRS ID.
        \STATE Radar sends out AoI (Area of Interest).
        \STATE Radar decodes the IRS reflection angle.
        \FOR {Each IRS reflection angle $\alpha$}
            \IF {$\alpha$ in AoI}
            \STATE Starts sensing using 2D-music.
            \ELSE
            \STATE Waiting for next $\alpha$.
            \ENDIF
        \ENDFOR
        \STATE Radar decides the AoI (Area of Interest) for next scanning round.
    \ELSE
        \STATE Repeat sending out FMCW chirps.
    \ENDIF
\ENDWHILE
\end{algorithmic}
\end{algorithm}

Algorithm~\ref{alg:radar_working} describes the \sn radar working procedure as a slave device. The radar continuously sends out FMCW chirps to probe the environment  (Line 1) and then receives reflected signals and processes them to detect OOK-modulated bits by the IRS (Lines 2-3). If the OOK modulated information from the IRS is detected, the radar decodes the IRS's ID (Lines 4-5). The radar then sends its AoI to the IRS and tries to decode the IRS reflection angles (Lines 6-7). For each reflection angle $\alpha$ used by the IRS, the radar checks whether it lies within the AoI. If $\alpha$ is in the radar's AoI, it starts sensing using advanced techniques such as 2D-MUSIC (Lines 7-10); otherwise, the radar waits for the next reflection angle (Lines 11-13). Based on the sensing results, the radar decides its AoI for the next scanning round (Lines 14-15). If no IRS modulation is detected, the radar continues transmitting FMCW chirps and waits for a response (Line 17-19). Note that the 2D-MUSIC in Line 10 provides the radar with the path length $D_{RS} + D_{ST}$ in Figure \ref{fig:nlos_layout}. Since the radar knows $D_{RS}$, it can determine $D_{ST}$. Therefore, the radar can calculate the position of the target using (\ref{eq:pos_x}) and (\ref{eq:pos_y}).

\begin{figure}[!h]
    \vspace{-3mm}
    \centering
    \includegraphics[width=0.3\textwidth]{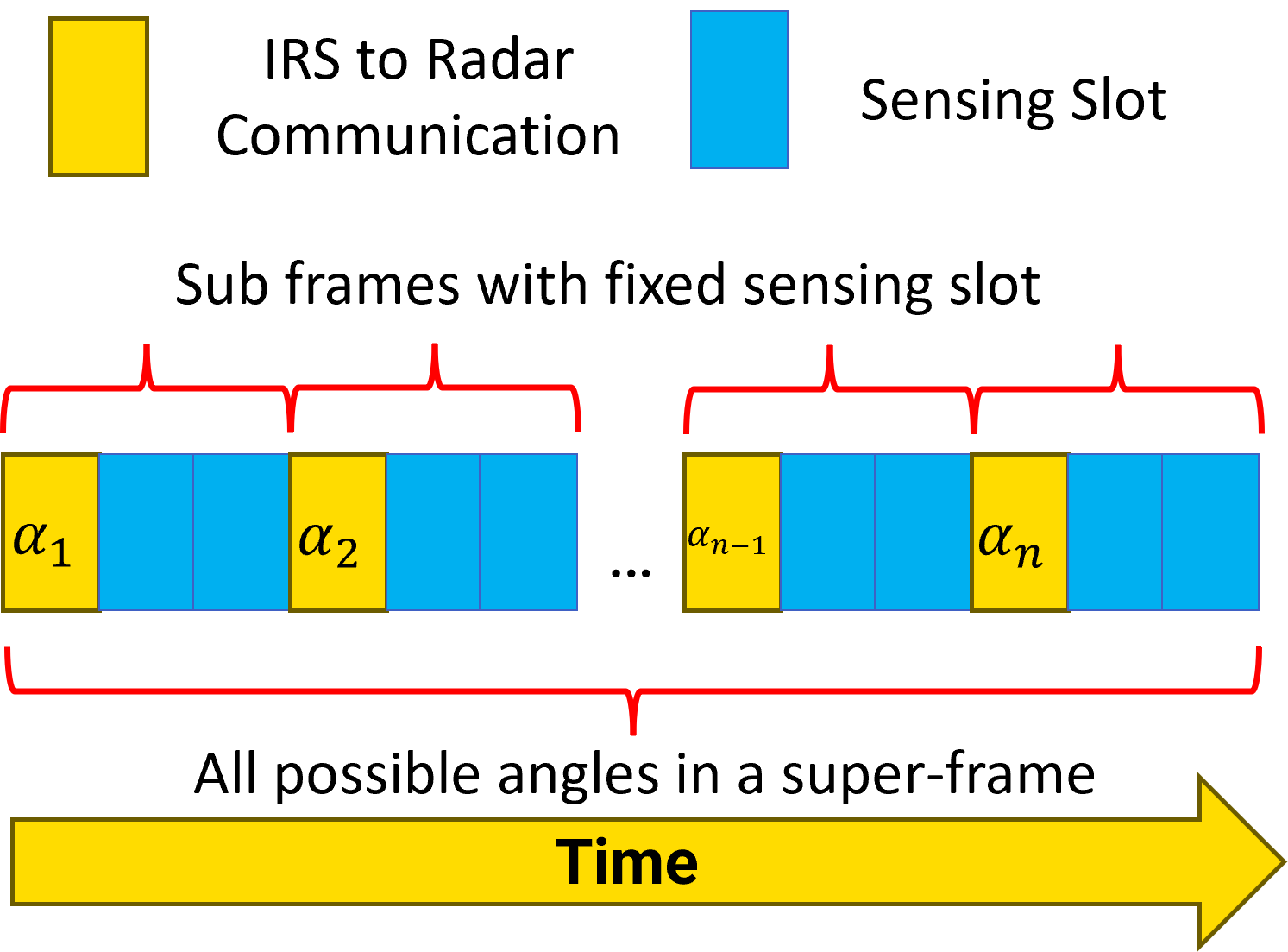}
    \caption{Naive Time Slots Allocation.}
    \label{fig:naive_allocation}
    \vspace{-3mm}
\end{figure}

\begin{figure}[!h]
    \vspace{-5mm}
    \centering
    \includegraphics[width=0.4\textwidth]{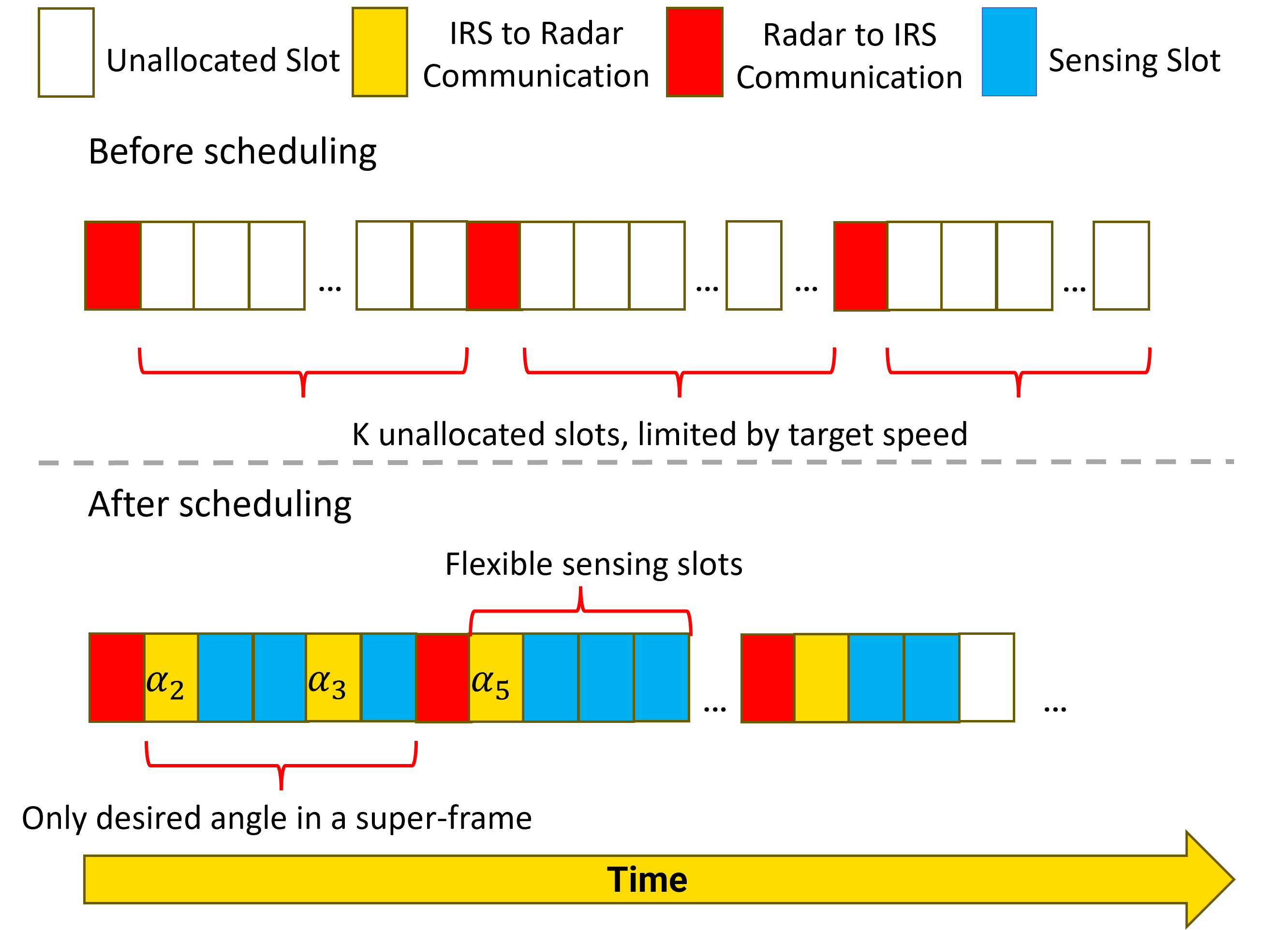}
    \caption{Time Series Slot Allocation Example.}
    \label{fig:optimized_allocation}
    \vspace{-5mm}
\end{figure}

\subsection{Adaptive Resource (Time Slot) Allocation}
\label{subsec:resourceAllocation}
Figure \ref{fig:naive_allocation} presents a naive approach to decide AoI. Here, we define a complete scan across all possible angles as a super-frame, with each angle corresponding to a sub-frame. Each sub-frame consists of time slots for communication (yellow) and sensing (blue). In a communication slot, the IRS operates in retro-reflection mode and broadcasts the reflection angle being used in the sub-frame. In the sensing slot, the IRS uses the reflection angle to allow the radar to localise the targets. 

Once the radar identifies the position of the IRS and receives the reflecting angle, it can localize targets in the NLoS area using the methods discussed in Section \ref{section:reflection_based_localization}.

\begin{algorithm} [htp]
    \caption{AoI-Sets construction}
    \label{alg:aoi_construct}
    \begin{algorithmic}[1]
        \STATE \textbf{Input:} $\mathcal{R}$: Raw Radar Data Matrix
        \STATE \textbf{\hspace{26pt}} $\mathcal{A}$: Previous scanning angle-duration set
        \STATE \textbf{\hspace{26pt}} $\Delta\alpha$: Scanning resolution in degrees
        \STATE \textbf{Output:} $\mathcal{T}$: Target Set
        \STATE \textbf{\hspace{33pt}} $\mathcal{P}$: Next scanning angle-duration set
        \FOR{each $A_i \in \mathcal{A}$}
        \STATE $\mathcal{T}\ \cup = $ MUSIC\_search($A_i$, $R$)
        \ENDFOR
        \IF{$\mathcal{T} = \phi$}
        \STATE $\mathcal{P} \gets \mathcal{A}$
        \STATE \textbf{Break}
        \ENDIF
        \STATE $T_{max\_energy} \gets $ max(Reflection\_energy\_search$(\mathcal{T}, R)$)
        \STATE $r_{max\_energy} \gets $ Range\_FFT $(T_{max\_energy}, R)$
        \STATE $D_{max\_energy} \gets $ Reflection\_duration\_search$(T_{max\_energy}, R)$
        \STATE  $r_{max\_speed}, v_{max} \gets$ max(Range\_Doppler$(\mathcal{T}, R)$)
        \FOR{$T_i \in \mathcal{T}$}
        \STATE $a_i \gets $ Reflection\_angle\_search($T_i$, $R$)
        \STATE $r_{i} \gets $ Range\_FFT $(T_{i}, R)$
        \ENDFOR
        \STATE $\underset{D_i}{\textnormal{minimise}}\quad D_{scan} = \sum_{i=1}^{len(\mathcal{T})}D_i$
        \STATE  $\textnormal{subject to} \quad$
        \STATE \textbf{\hspace{40pt}} $D_i \geq \frac{r_{max\_energy}^4}{r_i^4} D_{max\_energy}$
        \STATE \textbf{\hspace{40pt}} $v_{max} \times D_{scan} < \frac{\Delta\alpha}{180}\pi r_{max\_speed}$
        \FOR{each $a_i$}
        \STATE $D_i \gets $ Round($D_i$)
        \STATE $\mathcal{P}\ \cup = \{ (a_i,D_i) \}$
        \ENDFOR

    \end{algorithmic}
\end{algorithm}

By default, as is shown in Figure \ref{fig:naive_allocation}, the IRS needs to fully scan all angles to locate the targets. For a fast target, it might go undetected and move out of the detection area, potentially causing collisions if the scanning latency for each angle is too long. Moreover, during a full scan, a significant portion of the scanning time is wasted, as there may be no targets in some reflection angles. To address this issue, we introduced an extra time slot for radar-to-IRS communication (colored in red), shown in Figure \ref{fig:optimized_allocation}, between each super-frame, and the total number of time slots between two radar-to-IRS communication slots is limited based on the target's speed. This allows the radar to dynamically adjust the AoI for sensing. After the communication phase, the IRS will only select reflection angles based on the AoI set communicated to it from the radars, resulting in a reduced super-frame size and improved accuracy. This adaptive approach reduces localization time by focusing the scan on relevant areas. Additionally, 
for a given mmWave radar sensing environment with targets at different positions, the amount of allocated sensing chirps $N$ has a positive effect on SNR due to processing gain, which improve sensing, e.g., localization, accuracy. 

To determine the AoI sets and their corresponding allocated sensing slots, we begin with the default all-angle scanning that outputs the target sets and define the scanning resolution $\Delta\alpha$ to the gap between each scanning angle. As illustrated in Algorithm~\ref{alg:aoi_construct}, our approach essentially operates in a greedy manner: it first adds as many scanning angles as possible to form the initial AoI set \( U \), and then constructs the AoI set for the next scanning round.
The algorithm begins with a time-series  to help with monitoring new targets. After the full-angle scan, the algorithm gathers all target positions into the target set \( \mathcal{T} \) (Lines 6--8). Among these, it identifies the target with the highest reflection energy, \( T_{\text{max\_energy}} \), which determines the baseline for allocating time slots to any target (Lines 13--15). 
For targets with lower reflection energy, additional time slots (\( D_i \)) are allocated proportionally to their SNR relative to \( T_{\text{max\_energy}} \) (Line 23). Simultaneously, the algorithm identifies the target with the highest speed, \( v_{\text{max}} \), using the range-Doppler method (Line 16). This target's reflection must remain within the same angular resolution area, \( \frac{\Delta\alpha}{180}\pi r_{\text{max\_speed}} \), during the scan period constrained by \( v_{\text{max}} \times D_{\text{scan}} \) (Line 24). Here, \( D_{\text{scan}} \), the total scanning duration, is calculated as the sum of all allocated time slots (\( D_i \)) across the detected targets (Line 21).
With these constraints established, the algorithm employs a greedy approach to allocate time slots (\( D_i \)) to each target (Lines 17--24). After all targets are incorporated into the angle-duration set (AoI set), the algorithm updates \( \mathcal{P} \) (Lines 25--27). The system then pauses to collect new radar data and constructs updated AoI sets based on the latest target positions.

\section{Performance Evaluation}

\subsection{\sn Prototyping}
\begin{figure}[!htp]
    \vspace{-0.3cm}
    \centering
    \includegraphics[width=0.3\textwidth]{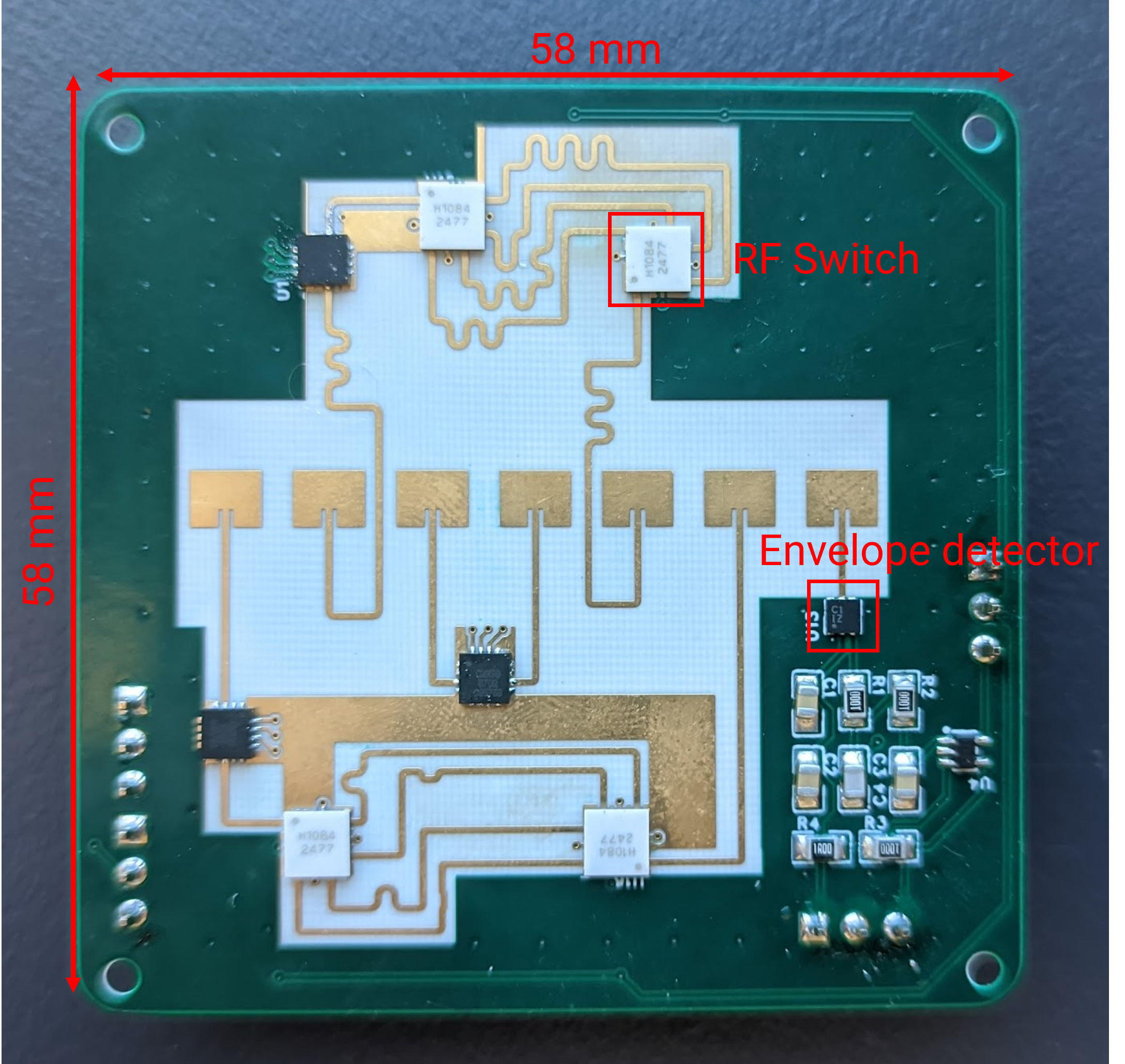}
    \caption{\sn 24 GHz IRS prototype with dimensions \( 58 \times 58 \, \mathrm{mm}^2 \).}
    \label{fig:pcb}
        \vspace{-0.3cm}
\end{figure}

\begin{figure}[!htp]
    \vspace{-3mm}
    \centering
    \includegraphics[width=0.45\textwidth]{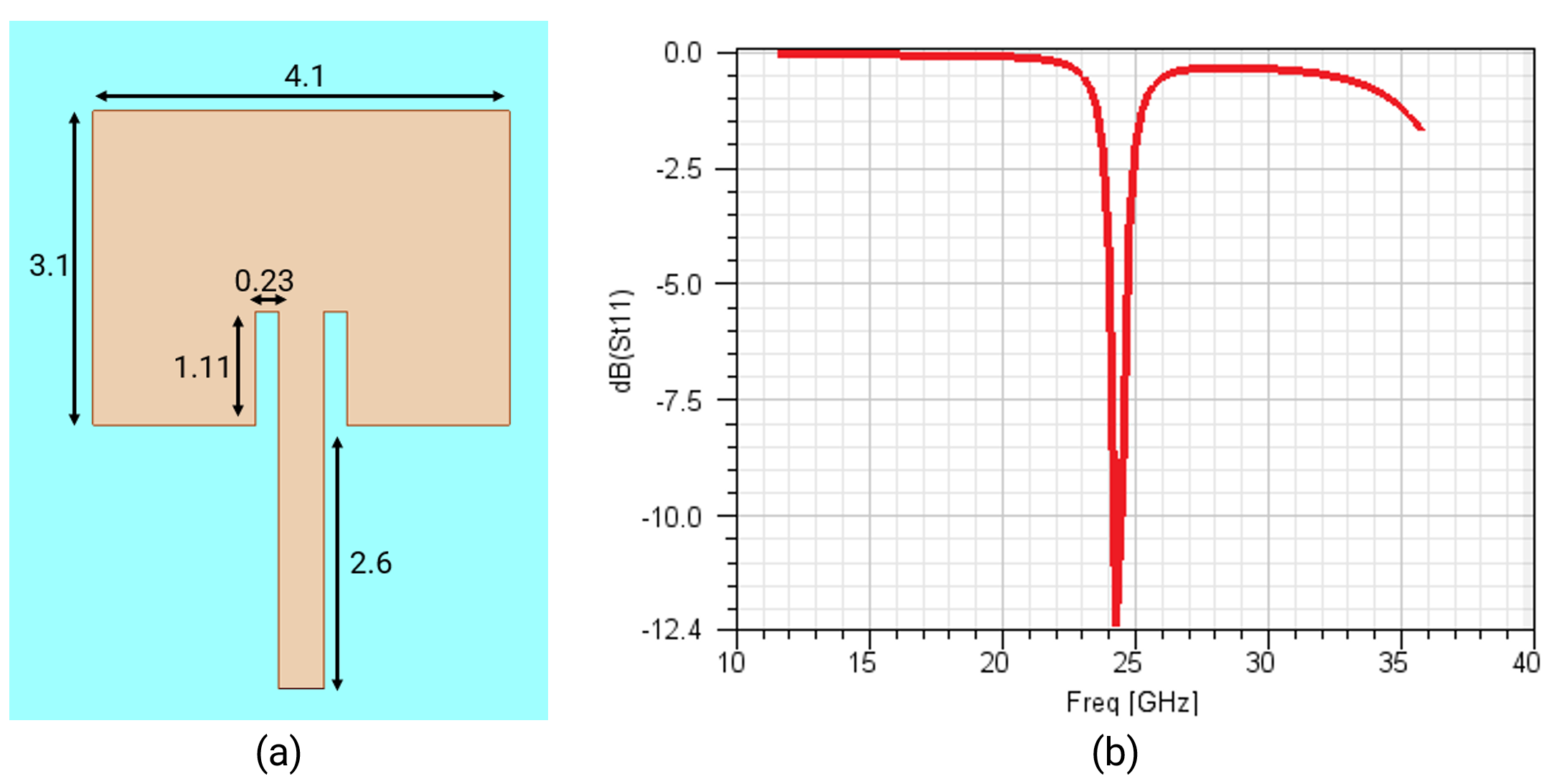}
    \caption{(a) \sn 24 GHz IRS antenna design. (b) Antenna performance in terms of S11.}
    \label{fig:antenna_design}
        \vspace{-0.3cm}
\end{figure}

Figure~\ref{fig:pcb} illustrates the \sn IRS prototype, with dimensions of \( 58 \times 58 \, \mathrm{mm}^2 \), fabricated using Rogers RO4003C radio frequency PCB. The IRS is controlled by an ATmega328P MCU~\cite{mcu}, and HMC1084LC4TR RF switches~\cite{switch} are employed for OOK modulation and transmission line selection to achieve retro-reflection or the desired reflection angle.
As shown in Figure~\ref{fig:antenna_design}, we designed insert-fed patch antennas with a peak gain of approximately 11\,\(\mathrm{dB}\) and an S11 below \(-10 \, \mathrm{dB}\) over the frequency range of 24–24.25\,\(\mathrm{GHz}\). These antennas connect to RFC ports, with four transmission lines featuring varying delays, enabling four distinct reflection angles and forming a Non-Retro-Reflection VAA.

An envelope detector (Analog Devices ADL6010)~\cite{envelope} is utilized to demodulate radar-to-IRS communication signals. The \sn system was implemented and evaluated using the EV-TINYRAD24G (Analog Devices)~\cite{radar}, a COTS 24\,\(\mathrm{GHz}\) radar with a bandwidth of 250\,\(\mathrm{MHz}\). This radar is equipped with two transmit and four receive antennas.

\subsection{Evaluation Setup}

\begin{figure}[htbp]
    
    \includegraphics[width=0.48\textwidth]{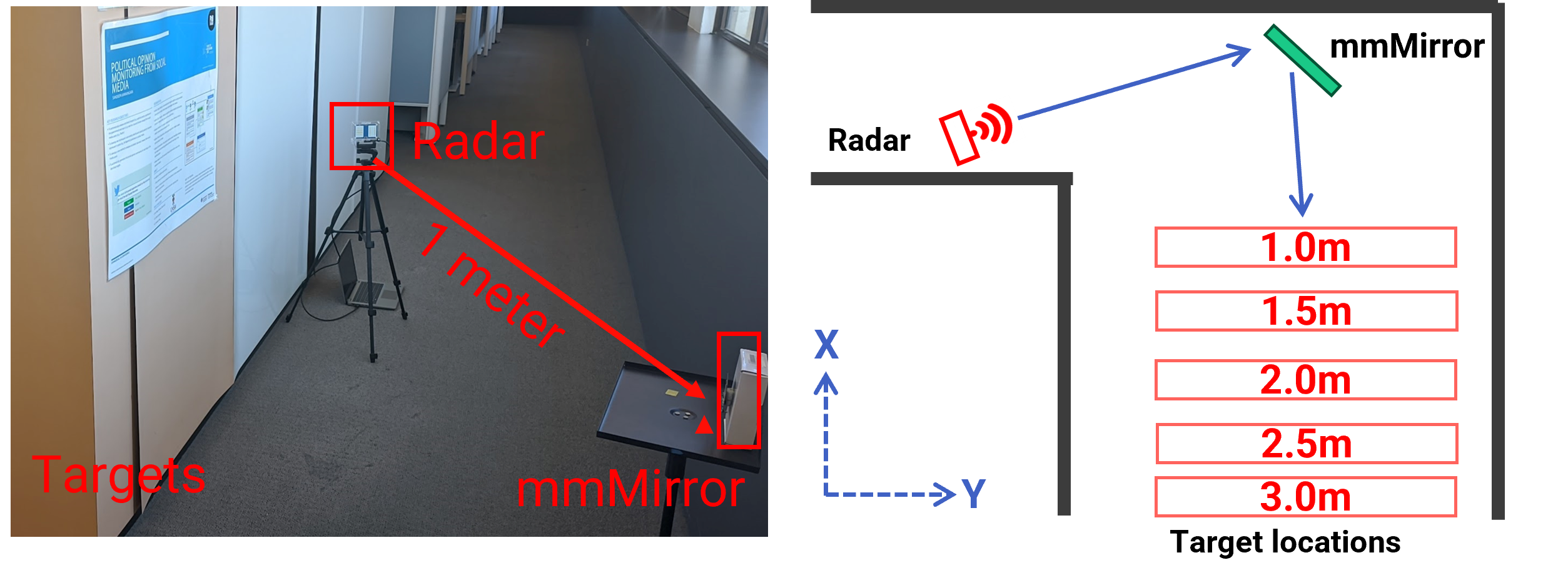}
    \centerline{(a)}
	\\
    \vspace{2mm}
    \includegraphics[width=0.47\textwidth]{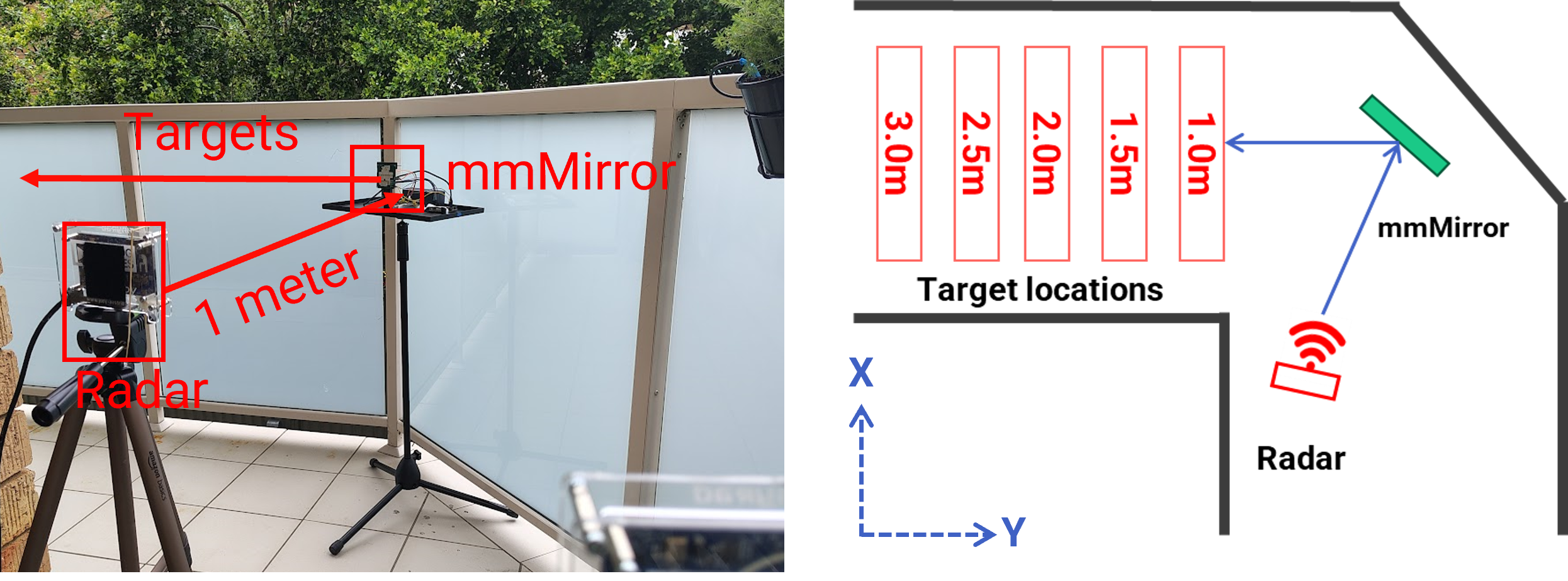}
    \centerline{(b)}
    \\
    \vspace{2mm}
    \includegraphics[width=0.48\textwidth]{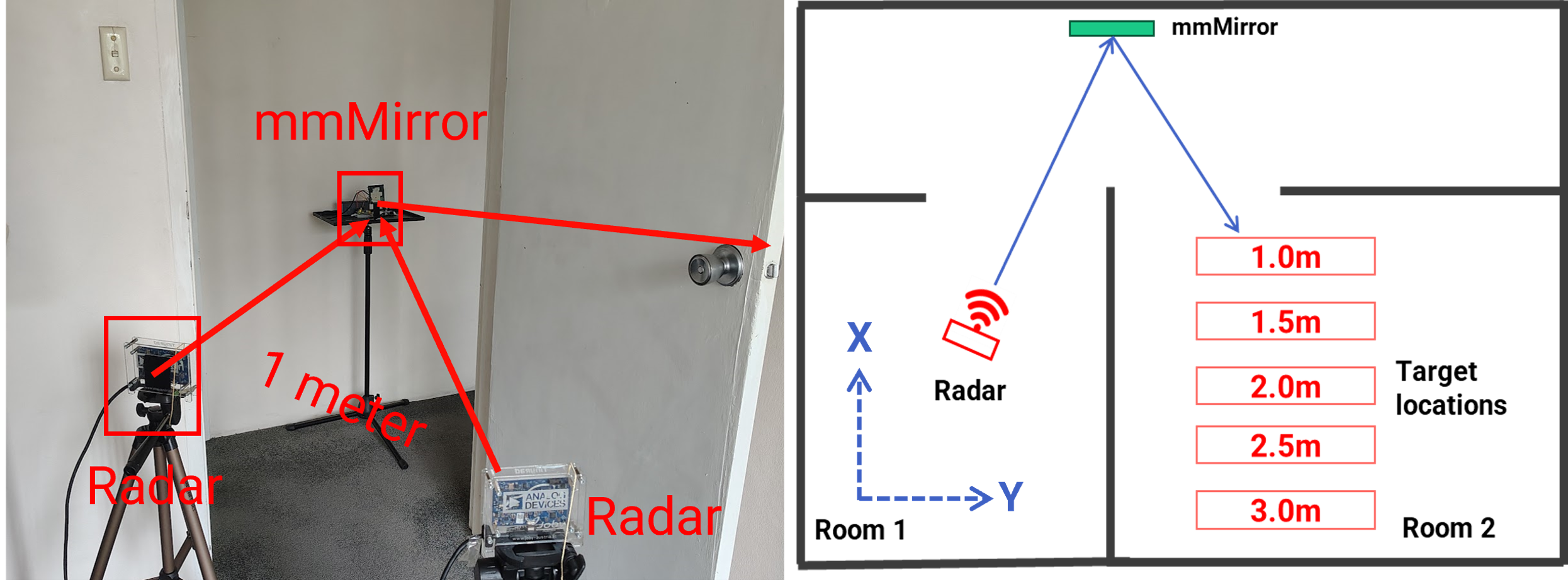}
    \centerline{(c)}
    \vspace{-5mm}
    \caption{\sn L-shaped corridor setup in 3 different environments, with targets positioned at 1\,m, 1.5\,m, 2\,m, 2.5\,m, and 3\,m from the IRS.}
    \label{fig:setup}
    \vspace{-3mm}
\end{figure}
\sn was deployed in three types of L-shaped corridors for performance evaluation: (a) Lab, (b) Balcony, and (c) Twin rooms, as shown in Fig.~\ref{fig:setup}. In each setup, one or two radars were positioned on one side of the \sn, while targets were placed on the opposite side, outside the radar’s field of view, creating a non-penetrable NLoS scenario. The radars remained fixed, meaning the robot halted upon detecting an IRS, similar to a traffic ``STOP'' sign.  

We evaluated communication performance across various ranges and angles, with radar-IRS distances spanning 0.5--2.0\,m and angles ranging from 0$^\circ$ to 60$^\circ$. The maximum achievable data rate was constrained by the antenna switching frequency ($F$) and the MCU’s sampling rate. For low-power radar-to-IRS communication, $F$ was set to 10\,Hz in single-radar scenarios and 5\,Hz or 10\,Hz in multi-radar setups.  

To assess target localization at different distances, we selected four target positions aligned with the \sn beam angle. Experiments are conducted with both \textit{single} and \textit{multiple} human targets moving randomly between points, either approaching or receding at speeds of 0.5--1\,m/s. The evaluated IRS-target distances are 1.0\,m, 1.5\,m, 2.0\,m, 2.5\,m, and 3.0\,m, with the radar positioned 1.0\,m from the IRS. The total radar-target distance are 2.0\,m, 2.5\,m, 3.0\,m, 3.5\,m, and 4.0\,m. Localization performance was measured based on errors along the x and y axes, as well as overall distance errors, all reported in centimeters.  

We limited the maximum IRS-target distance to 3\,m due to mmWave signal attenuation and IRS reflection loss. As multipath effects intensify when a target moves deeper into the corridor, reflections become more prominent, though the immediate collision risk decreases. \sn is designed to prioritize the closest target, as it poses the highest risk of collision.


\subsection{Communication Performance}
\begin{figure}[htbp]
    \vspace{-4mm}
    \centering
    \begin{minipage}[t]{0.5\linewidth}
        \includegraphics[width=1\textwidth]{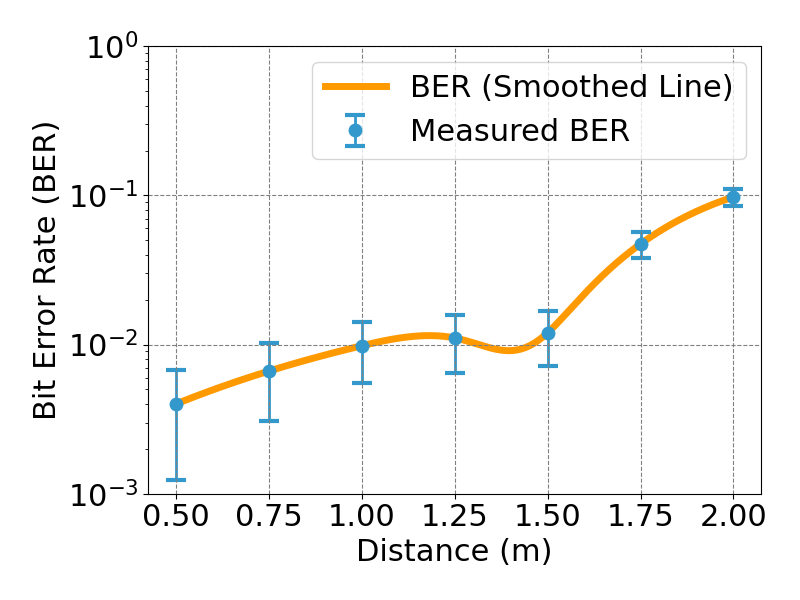}
        \centerline{(a)}   
    \end{minipage}%
    \begin{minipage}[t]{0.5\linewidth}
        \includegraphics[width=1\textwidth]{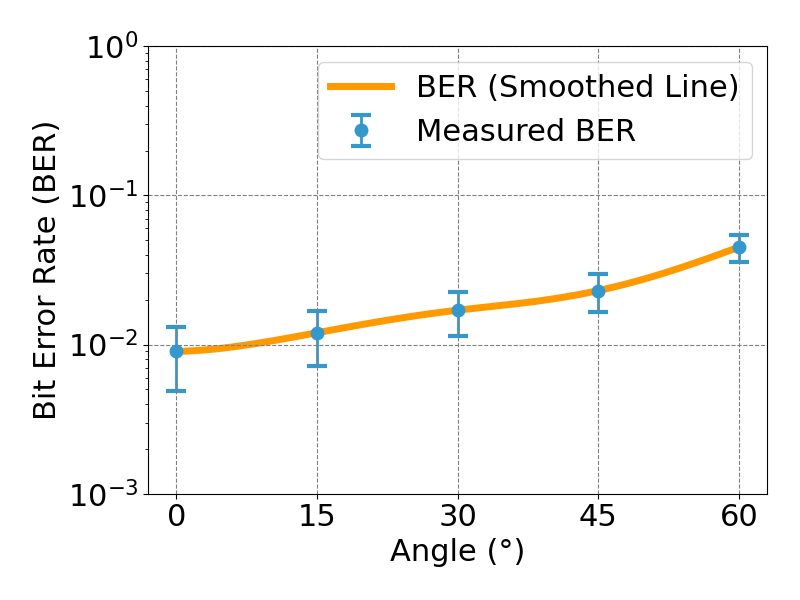}
        \centerline{(b)}
    \end{minipage}
    \caption{Radar-to-IRS communication BER vs (a) distance and (b) angle, with 95\% confidence intervals.}
    \label{fig:ber}
    \vspace{-4mm}
\end{figure}

We evaluated \sn’s radar-to-IRS communication under varying distances and angles. Figure~\ref{fig:ber}(a) presents the bit error rate (BER) as a function of radar-to-IRS distance. As the distance increases from 0.5\,m to 2.0\,m, the BER rises due to the radar’s limited transmission power and the sensitivity of the envelope detector. Nonetheless, the system achieves a BER below \(10^{-1}\) at 2.0\,m, which is sufficient for accurate localization in indoor NLoS scenarios where moving objects, such as robots and humans, typically have low velocities (e.g., 1\,m/s). These results indicate that a radar-equipped device, such as a robot, can effectively communicate with an IRS deployed at a corner 2\,m away and take appropriate actions based on the received information.

Figure~\ref{fig:ber}(b) shows the BER as a function of radar-to-IRS angles. The results reveal that the BER increases with larger angles (0°–60°), reflecting the impact of orientation on communication performance. Despite this, the BER remains below \(10^{-1}\) even at 60°, demonstrating robust communication capabilities in typical indoor environments such as corridors.

\subsection{Localization Performance}

\begin{figure}[htbp]
    \vspace{-4.8mm}
    \centering
    \begin{minipage}[t]{0.5\linewidth}
        \includegraphics[width=1\textwidth]{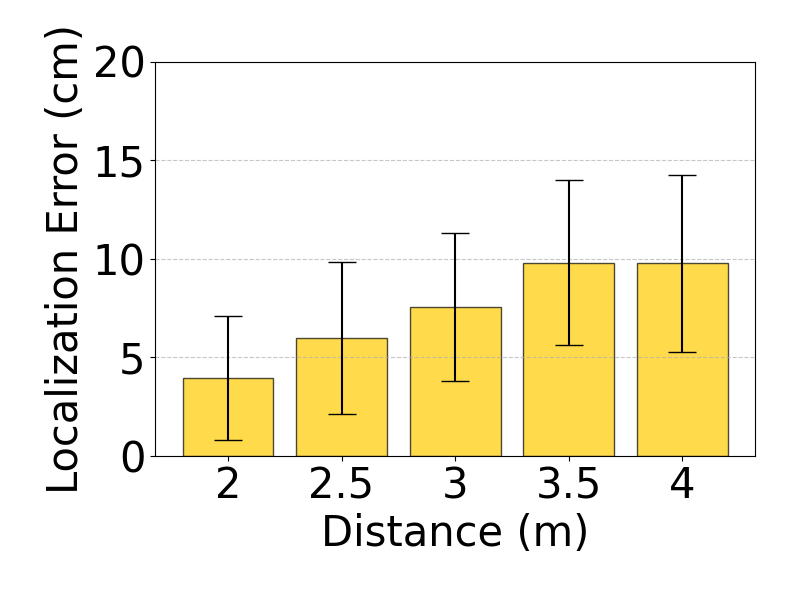}
        \centerline{(a)}   
    \end{minipage}%
    \begin{minipage}[t]{0.5\linewidth}
        \includegraphics[width=1\textwidth]{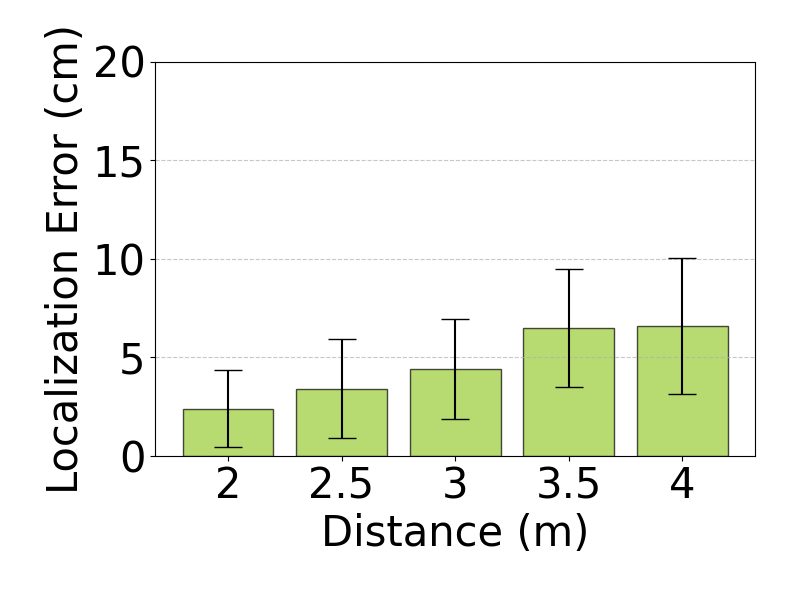}
        \centerline{(b)}
    \end{minipage}
    \caption{Median single-target localization errors along the (a) x-axis, (b) y-axis.}
    \label{fig:error_xy_single}
    \vspace{-4mm}
\end{figure}


\textbf{Single-Target Performance:} Figure~\ref{fig:error_xy_single} illustrates \sn’s localization performance in terms of distance errors for the single-target setup along the x and y axes at various target distances. The median localization errors along the x-axis are 3.96\,cm at 2\,m, 5.94\,cm at 2.5\,m, 7.82\,cm at 3\,m, 9.87\,cm at 3.5\,m, and 9.96\,cm at 4\,m. Along the y-axis, the corresponding errors are 2.38\,cm, 3.40\,cm, 4.54\,cm, 6.69\,cm, and 6.75\,cm. The overall median localization errors are 4.63\,cm at 2\,m, 6.8\,cm at 2.5\,m, 8.61\,cm at 3\,m, 11.28\,cm at 3.5\,m, and 11.83\,cm at 4\,m. These results demonstrate that the proposed system achieves reliable localization accuracy in single-target scenarios.


\begin{figure}[htbp]
    \vspace{-5mm}
    \centering
    \begin{minipage}[t]{0.5\linewidth}
        \includegraphics[width=1\textwidth]{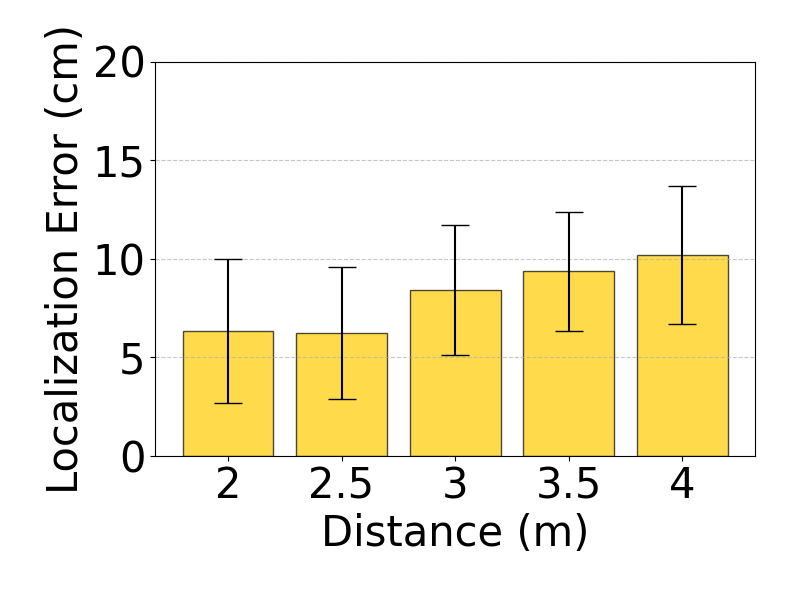}
        \centerline{(a)}   
    \end{minipage}%
    \begin{minipage}[t]{0.5\linewidth}
        \includegraphics[width=1\textwidth]{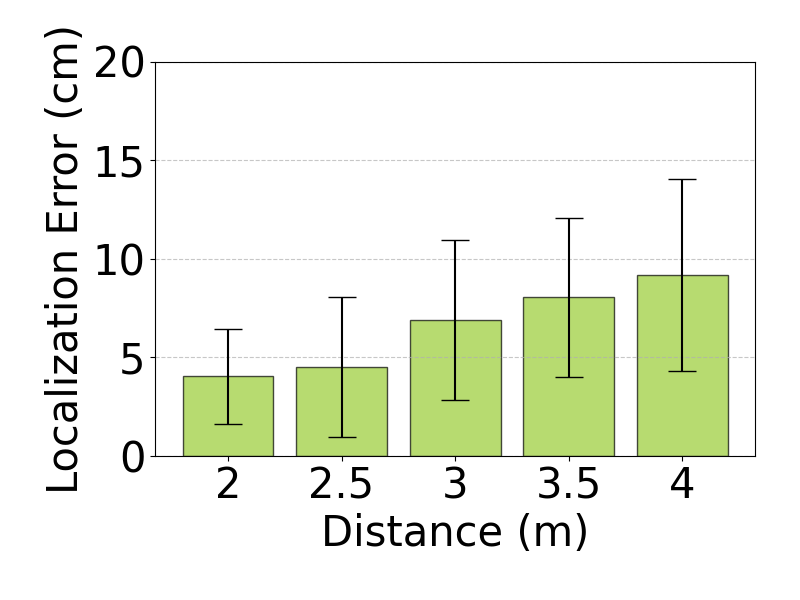}
        \centerline{(b)}
    \end{minipage}
    \caption{Median multiple-target localization errors along the (a) x-axis and (b) y-axis.}
    \label{fig:error_xy_multi}
    \vspace{-4mm}
\end{figure}

\textbf{Multiple-Target Performance:} Figure~\ref{fig:error_xy_multi} presents \sn’s localization performance along the x and y axes, as well as the overall distance errors at various target distances. The system demonstrates robust accuracy using a single IRS. The median localization errors along the x-axis are 6.31\,cm at 2\,m, 6.28\,cm at 2.5\,m, 8.46\,cm at 3\,m, 9.42\,cm at 3.5\,m, and 10.18\,cm at 4\,m. Along the y-axis, the corresponding errors are 4.10\,cm, 4.63\,cm, 6.89\,cm, 8.12\,cm, and 9.17\,cm. The median overall distance errors are 7.49\,cm at 2\,m, 7.70\,cm at 2.5\,m, 10.89\,cm at 3\,m, 12.31\,cm at 3.5\,m, and 13.68\,cm at 4\,m. These results highlight the system’s ability to maintain high localization accuracy even in multi-target scenarios.

\begin{figure*}[htbp]
    \vspace{-5mm}
    \begin{minipage}[t]{0.33\linewidth}
        \centering
        \includegraphics[width=1\textwidth]{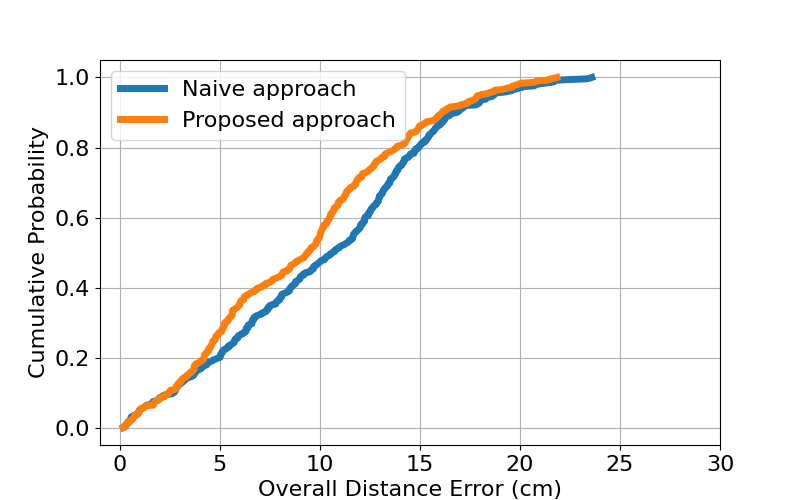}
        \centerline{(a)}
    \end{minipage}%
    \begin{minipage}[t]{0.33\linewidth}
        \centering
        \includegraphics[width=1\textwidth]{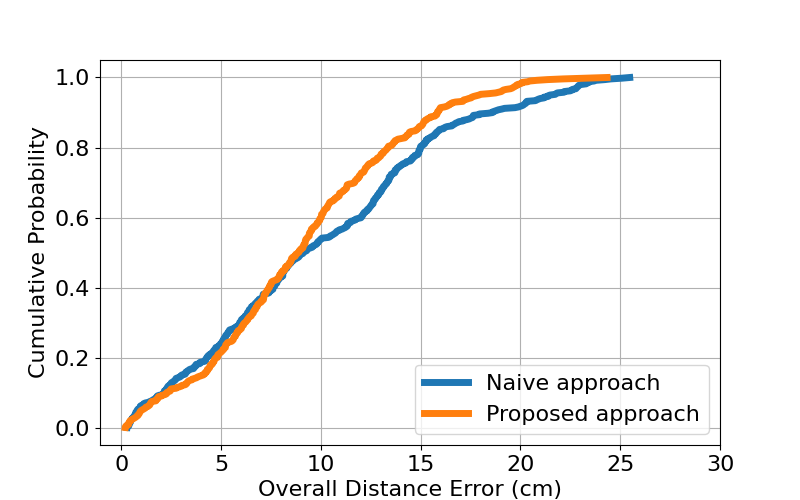}
        \centerline{(b)}
    \end{minipage}%
    \begin{minipage}[t]{0.33\linewidth}
        \centering
        \includegraphics[width=1\textwidth]{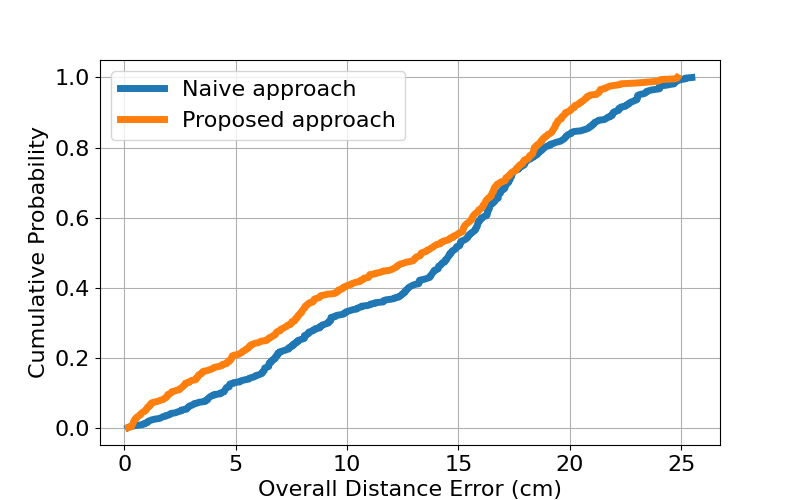}
        \centerline{(c)}
    \end{minipage}%
    \caption{CDFs for localization errors in (a) Lab, (b) Balcony, and (c) Twin rooms.}
     \vspace{-0.2cm}
    \label{fig:error_multi_env}
    \vspace{-0.3cm}
\end{figure*}

To assess the robustness of the proposed \sn system, Figure~\ref{fig:error_multi_env} shows the cumulative distribution functions (CDFs) of localization errors across different indoor environments, including Lab, Balcony, and Twin rooms. The results reveal two key findings. First, \sn achieves consistently good localization accuracy in all three environments, with the \(50^{\text{th}}\) percentile localization errors measured at 7.89\,cm for Lab, 8.75\,cm for Balcony, and 12.75\,cm for Twin rooms. Second, our proposed method outperforms the baseline approach, and further analysis will be presented in the next section.

\subsection{Key Algorithm Performance}

This section evaluates the impact of the adaptive time slot allocation algorithm introduced in Section~\ref{subsec:resourceAllocation} by comparing its performance against a baseline approach that performs a full scan of all IRS angles. The evaluation includes both single-target and multiple-target scenarios.

\begin{table}[ht]
\centering
\caption{Scanning Time (s) for Different Scenarios}
\vspace{-4mm}
\begin{tabular}{|l|c|c|}
\hline
\textbf{Scenario}         & \textbf{Average}  & \textbf{\(95^{\text{th}}\) Percentile} \\ \hline
Baseline                  & 7.15                                & -  \\ \hline
Multiple targets          & 4.13                                & 5.96  \\ \hline
Single target             & 1.80                                & 2.77  \\ \hline
\end{tabular}
\label{tab:scanning_time}
\vspace{-4mm}
\end{table}

\begin{figure}[htbp]
    \vspace{-0.45cm}
    \begin{minipage}[t]{0.5\linewidth}
        \centering
        \includegraphics[width=1\textwidth]{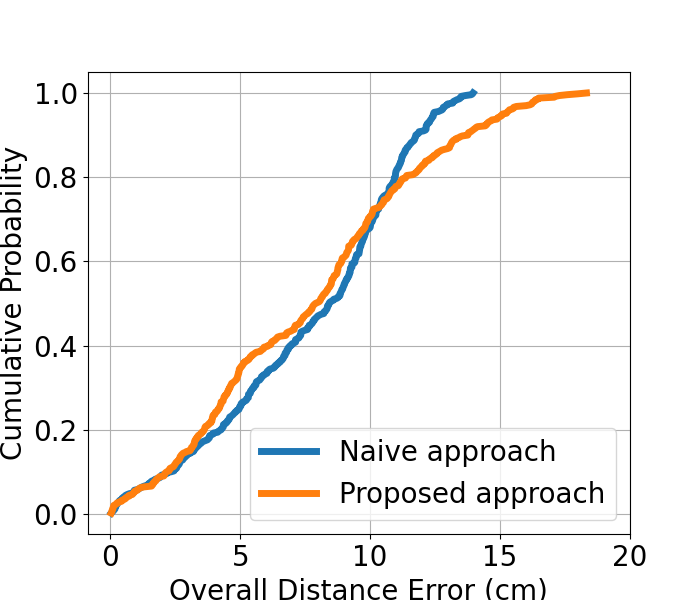}
        \centerline{(a)}
    \end{minipage}%
    \begin{minipage}[t]{0.5\linewidth}
        \centering
        \includegraphics[width=1\textwidth]{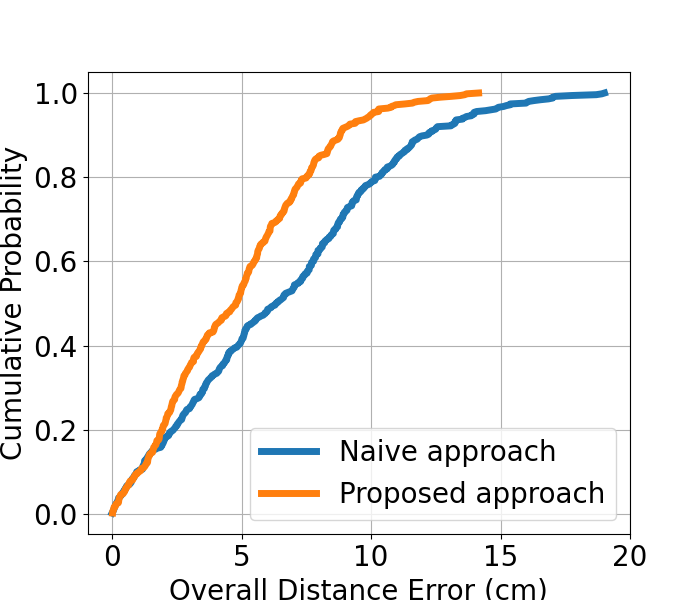}
        \centerline{(b)}
    \end{minipage}%
   \vspace{-0.2cm}
    \caption{CDFs for localization errors along the (a) x-axis, (b) y-axis.}
    \label{fig:cdf_xy}
    \vspace{-0.3cm}
\end{figure}

Table~\ref{tab:scanning_time} summarizes the average scanning times and \(95^{\text{th}}\) percentile scanning time required for the IRS to complete a full scan under different scenarios. The localization results for these approaches, represented as cumulative distribution functions (CDFs), are shown in Figure~\ref{fig:cdf_xy}. The proposed time slot allocation algorithm significantly outperforms the baseline full-scan scheme. Notably, it reduces the average scanning time from 7.15\,s to 4.13\,s and the \(95^{\text{th}}\) Percentile Scanning Time (s) to 5.96\,s in the multiple-target scenario and further to 1.80\,s and 2.77\,s in the single-target scenario. Consider a potential robot-human collision scenario \cite{vslajpah2021effect} where the robot detects and stop at 2\,m from the IRS, while a human, walking at a speed of 0.8\,m/s, approaches from the other side at a distance of 3\,m. The total reaction time is 6.25\,s, providing sufficient time for the robot to detect the human and take action to avoid a collision. Moreover, the sensing time can be further reduced by allocating smaller sensing time slots (i.e., using fewer chirps per slot). However, this introduces a trade-off between accuracy and latency.


The proposed allocation algorithm achieves an \(50^{\text{th}}\) percentile localization error of 7.89\,cm along the x-axis and 4.84\,cm along the y-axis, compared to 8.34\,cm and 6.49\,cm, respectively, using the baseline approach. This improvement arises from the algorithm's ability to discard reflection angles without targets and reallocate unused sensing time to angles associated with weaker reflections. 

The observed performance improvements are attributed to the reconfigurable IRS design and the time slot allocation algorithm introduced in Section~\ref{subsec:resourceAllocation}. However, it is important to note that our approach may occasionally underperform compared to the baseline method in low-SNR scenarios. This is because, unlike the naive approach—which uniformly allocates excessive time to all directions—our method dynamically adjusts time slots based on specific directions. Consequently, the total sensing time may be lower than that of the naive approach, leading to reduced performance in certain cases. Nevertheless, as previously discussed, \sn is designed to prioritize closer, high-SNR targets, as they pose a greater risk of collision.

\subsection{Power Consumption}

We measured the power consumption of \sn using the Power Profiler Kit II \cite{power_profiler} from Nordic Semiconductor. The primary components of \sn include RF switches, an envelope detector, and an MCU. The system’s power consumption is categorized into three operating modes:

\begin{enumerate}
    \item \textbf{Sleeping Mode}:  
    In this mode, the IRS remains idle, continuously monitoring for incoming FMCW signals, as no radar is approaching the \sn. To minimize power usage, the MCU operates in a low-power sleep state, with all other components turned off. Under these conditions, the typical power consumption of the MCU is approximately 46.5\,\(\mu\)W.

    \item \textbf{Communicating Mode}:  
    In this mode, both the envelope detector and the MCU are active, consuming approximately 8\,mW and 641.058\,\(\mu\)W, respectively. To optimize energy efficiency, the MCU is periodically disabled using a hardware timer running at 64\,Hz. This approach enables the MCU to wake intermittently to power cycle the envelope detector, perform ADC sampling, and decode signals. As a result, the total average power consumption in this mode is approximately 2.763\,mW.

    \item \textbf{Reflection Mode}:  
    During reflection mode, both the RF switch and the MCU are active. Each RF switch consumes approximately 2.86\,\(\mu\)W. By applying the same periodic wake-up strategy, components are powered only as needed, reducing the total average power consumption in this mode to approximately 50\,\(\mu\)W.
\end{enumerate}

To evaluate practical energy usage, we consider a scenario where \sn spends 50\% of its time in idle (sleeping mode) and 50\% in communicating and reflecting modes. Assuming that 10\% of the time slots are allocated for radar-to-IRS communication, the total power consumption of the tag is estimated at 183.9\,\(\mu\)W. With an AA battery rated at 2,500\,mAh, \sn is projected to operate continuously for approximately 566 days.

\section{Discussion and Future Work}

\noindent \textbf{IRS Location:}  
The placement of the IRS significantly affects localization accuracy due to antenna gain patterns and VAA reflection characteristics. Since this study employs a 2D-VAA design, any misalignment between the IRS and radar antennas can weaken reflected signals, increasing localization errors. To ensure optimal reflection energy, the IRS and radar are assumed to be at the same height. Future work could explore enhancements such as deploying multiple VAA arrays to optimize reflections at varying heights or adopting 3D-reflection VAA arrays, as proposed in recent studies. While promising, these approaches fall beyond the scope of this work.  

\noindent \textbf{Number and Position of Targets:}  
The number and spatial arrangement of targets are crucial for effective device-free localization in NLoS scenarios. When multiple targets fall within the same reflection angle, traditional 2D-MUSIC algorithms may struggle to differentiate them from multipath components, degrading localization performance. The current \sn system detects multiple targets only if they occupy distinct beam-scanning angles, with at most one target per angle. This design makes \sn well-suited for human presence detection in indoor environments, even with a high target density.  
To mitigate challenges from multiple targets within the same \sn scanning angle, future work could explore increasing the number of scanning angles or integrating advanced signal processing techniques, offering promising directions for further research.



\section{Related Work}

\noindent \textbf{Van Atta Arrays:}  
VAAs enable diverse applications across multiple domains. Millimetro~\cite{soltanaghaei2021millimetro} achieved centimeter-level localization using VAA-based retro-reflection, while Trzebiatowski et al.~\cite{trzebiatowski2022simple} introduced a 60\,GHz VAA tag for mmWave identification.  

Traditional VAAs focus on azimuth-plane retro-reflection, but Hawkeye~\cite{bae2023hawkeye} extended this to 77\,GHz for full 3D coverage. OmniScatter~\cite{bae2022omniscatter} enabled backscatter communication using controlled signal modulation. Additionally, Ang and Eleftheriades~\cite{ang2018passive} demonstrated retro-redirection via phase delay adjustments.  

However, most studies address LoS scenarios, where VAAs perform reliably. Their effectiveness significantly degrades in fully NLoS conditions, limiting their utility in obstructed environments.  

\noindent \textbf{IRS-Aided Communication and Localization:}  
IRSes enhance reflected signal strength and simplify propagation models~\cite{liu2022sensing, lau2012reconfigurable, wu2019towards, ye2020joint}, with applications in both positioning and communication. Song et al.~\cite{song2022intelligent} used an IRS-assisted link for NLoS DoA estimation, while Wen et al.~\cite{wen2024polarized} introduced a polarized IRS architecture that eliminates prior channel knowledge. Esmaeilbeig et al.~\cite{esmaeilbeig2022irs} demonstrated that multiple IRSes improve radar performance.  

Although IRS studies have advanced localization, most focus on theoretical models and LoS conditions. Dodds et al.~\cite{Dodds2024Around} explored outdoor NLoS localization via mmWave radar but relied on additional equipment like LiDAR or tagged targets, limiting real-world scalability.

\noindent \textbf{Contributions of \sn:}  
In contrast to existing works, \sn achieves accurate NLoS localization without relying on tagged targets or extra equipment (e.g., LiDAR). Our system integrates a VAA-based IRS capable of supporting multiple reflection angles, a novel antenna encoding scheme to enable seamless radar-IRS communication, and a time slot allocation algorithm that dynamically adjusts reflection angles and sensing duration. These contributions enhance SNR, improve localization accuracy, and reduce latency, demonstrating substantial potential for deployment in complex indoor environments.

\section{Conclusion}

This paper presents \sn, a novel mmWave NLoS sensing system designed for precise, device-free indoor localization using a single radar and an  IRS. The system leverages a VAA-based reconfigurable IRS and integrates with existing FMCW signals. A key innovation of \sn is its antenna modulation scheme, which enables efficient radar-IRS communication, dynamic reflection angle adjustments, and effective differentiation of the IRS from static reflectors. By utilizing a frequency-based antenna encoding scheme and time slot allocation algorithm, the system effectively handles signal separation, improves SNR, and reduces sensing latency and support multiple radars and multiple targets scenarios.
Extensive evaluations in diverse indoor environments demonstrate that \sn achieves high accuracy in device-free localization, even under challenging NLoS conditions. These results highlight the system’s substantial potential for real-world applications, particularly in enabling safe and efficient robot-human interaction in complex indoor environments.




\balance
\bibliographystyle{ACM-Reference-Format}
\bibliography{references.bib}

\end{document}